\documentclass[]{aa}
\usepackage{amsmath}
\usepackage{graphicx}
\usepackage{txfonts}
\usepackage{color}
\usepackage[percent]{overpic}
\usepackage{subfig}
\AtBeginDocument{\mathcode`v=\varv}

\begin{document}

\title{On the orbital evolution of binaries with circumbinary discs}

\author{R. M. Heath
  \and
  C. J. Nixon
}

\institute{\centering School of Physics and Astronomy, University of Leicester, Leicester, LE1 7RH, UK\\
}

\date{\today}


\abstract{Circumbinary discs are generally thought to take up angular momentum and energy from the binary orbit over time through gravitational torques mediated by orbital resonances. This process leads to the shrinkage of the binary orbit over time, and is important in a variety of astrophysical contexts including the orbital evolution of stellar binaries, the migration of planets in protoplanetary discs, and the evolution of black hole binaries (stellar and supermassive). The merger of compact object binaries provides a source of gravitational waves in the Universe. Recently, several groups have reported numerical simulations of circumbinary discs that yield the opposite result, finding that the binary expands with time. Here we argue that this result is primarily due to the choice of simulation parameters, made for numerical reasons, which differ from realistic disc parameters in many cases. We provide physical arguments, and then demonstrate with 3D hydrodynamical simulations, that thick (high pressure, high viscosity) discs drive sufficient accretion of high angular momentum material to force binary expansion, while in the more realistic case of thin (low pressure, low viscosity) discs there is less accretion and the binary shrinks. In the latter case, tides, which generally transfer angular momentum and energy from the more rapidly rotating object (the binary) to the less rapidly rotating object (the disc), are the dominant driver of disc-binary evolution. This causes the binary to shrink. We therefore conclude that for common circumbinary disc parameters, binaries with non-extreme mass ratios are expected to shrink over time. Expansion of the binary can occur if the disc viscosity is unusually high, which may occur in the very thick discs encountered in e.g. circumplanetary discs, super-Eddington AGN, or the outer regions of passive protostellar discs that are heated by the central protostar. We also provide discussion of the impact that some simplifications to the problem, that are prevalent in the literature and made usually for numerical convenience, have on the disc-binary evolution.}


\maketitle

\section{Introduction}
\label{intro}
Circumbinary discs are discs of matter that orbit externally to a central binary system that is typically composed of two stars or black holes. They may form in a variety of astrophysical systems, including when stellar binaries capture material in dense star forming regions or in galactic centres where, following a galaxy merger, two supermassive black holes (SMBH) accrete gas from the host galaxy.

In the standard picture \citep[see, for example,][]{Lin:1986aa,Pringle:1991aa,Artymowicz:1994aa}, a disc of gas starting at large radius from the binary evolves initially due to viscosity; the matter spreads to smaller and larger radii, facilitated by viscous torques that usually arise due to disc turbulence \citep{Shakura:1973aa}. Once the disc extends inwards to radii of order a few times the binary separation, the disc orbits resonate with the binary orbit at discrete locations in the disc \citep{Papaloizou:1977aa} resulting in angular momentum and energy being transferred from the binary to the disc. This, and other processes, can truncate the disc and prevent accretion of matter on to the binary orbit, and thus the disc resembles a decretion disc \citep{Pringle:1991aa}. The numerical simulations of \cite{Artymowicz:1994aa,Artymowicz:1996aa} showed that even for a truncated disc, matter could leak from the inner disc through time dependent streams that feed the binary. These original and important papers form the standard paradigm for external accretion on to binary systems, with the net effect resulting in the binary orbit decaying with time.

Recently, several groups have challenged this standard picture \citep{Miranda:2017aa,Tang:2017aa,Munoz:2019aa,Moody:2019aa,Munoz:2020aa}. They present 2D numerical hydrodynamic simulations of circumbinary discs with the binary orbit modelled as fixed \citep[][provide complementary 3D simulations for comparison]{Moody:2019aa}. Through detailed investigations of the torques acting between the disc and the binary in their simulations, these works find that the net torque on the binary is positive and that the specific angular momentum of the binary increases with time. Thus, they conclude that interaction with circumbinary discs causes the binary to expand with time -- and this conclusion has at times been stated without reference to the parameters of the disc-binary system. This conclusion is in stark contrast to the standard picture of circumbinary disc evolution.

Very recently, \cite{Tiede:2020aa} have challenged this conclusion drawn from these recent simulations. \cite{Tiede:2020aa} report numerical simulations, employing 2D (vertically integrated) Eulerian hydrodynamics, of circumbinary discs and find that the results are parameter dependent with thinner discs causing the binary to shrink, while thicker discs cause the binary to expand. \cite{Tiede:2020aa} hold the kinematic viscosity constant between simulations (fixing the value of $\alpha(H/R)^2$, where $\alpha$ is the \citealt{Shakura:1973aa} dimensionless viscosity parameter and $H/R$ is the disc angular semi-thickness) and they then vary $H/R$ between simulations (and thus implicitly also varying the value of $\alpha$). They find that the critical value of $H/R$ dividing the binary evolution between contraction and expansion is $H/R \approx 0.04$, which they define as a Mach number of $R/H = 25$\footnote{Note that by this definition of the disc Mach number, one might expect that the disc regularly shocks. However, while the orbital speed is supersonic (i.e.\@ $H/R \equiv c_{\rm s}/v_\phi < 1$), in general the turbulent velocities \citep[cf.][]{Martin:2019aa} and the differential orbital velocity over a radial scale of order $H$ are at most transsonic.}. They point out that the discs in active galactic nuclei (AGN), and therefore the ones which are relevant to the structure of discs around SMBH binaries, typically satisfy $H/R < 0.04$, indicating that in this case the binary will shrink with time. As we will see below, the results we present here broadly agree with \cite{Tiede:2020aa}, but also indicate that the critical value of $H/R$ that divides binary expansion and binary contraction depends on other system parameters, and is likely to be larger than the value suggested by \cite{Tiede:2020aa}. 

Following \cite{Tiede:2020aa}, here we provide (see Section~\ref{simulations}) two 3D numerical simulations with different disc thicknesses, one of which shows binary expansion, and one of which shows binary contraction; we also provide some preliminary results at intermediate disc thickness in Section~\ref{conclusions}. We note that all of the investigations above that find binary expansion assume that the binary orbit is fixed, while here we model the binary as ``live'' and allow it to evolve through interactions with the gas disc. This method allows for the binary eccentricity, and thus the location and strength of resonances in the disc, to evolve self-consistently as the binary semi-major axis evolves. We discuss the physical reasons governing the evolution of the binary orbit and connect this with the properties of the binaries and discs in different astrophysical systems.

The structure of the paper is as follows. In Section~\ref{discbinary} we describe the physics of disc-binary interaction. In Section~\ref{simulations} we present our numerical simulations. In Section~\ref{discussion} we provide discussion, and we conclude in Section~\ref{conclusions}.

\section{Disc--binary interaction}
\label{discbinary}
The interaction of a disc with a binary system is an important topic in astrophysics that has received a lot of attention. Broadly speaking the different types of disc-binary system can be split into two, one where the disc orbits primarily around one object with the second component of the binary providing an (external) perturbation \citep[e.g.\@][]{Papaloizou:1977aa,Goldreich:1979aa}, and the other where the disc orbits externally to both components of the binary (i.e.\@ a circumbinary disc)\footnote{A third type of system, which is essentially a hybrid of these two, occurs for very low mass ratio binaries, where the lower mass object orbits within the disc of the primary object. In this case, from the point of view of the secondary object, there is an internal and an external accretion disc. This commonly occurs for planets in circumstellar discs, and stars in AGN discs.}. While the discs in these two cases evolve differently, much of the physics of the interaction is the same \citep[see e.g.\@][]{Artymowicz:1994aa}. In this paper we are only concerned with the circumbinary disc case, although we note that accretion from a circumbinary disc can produce discs orbiting around each of the binary components, and conversely discs around each binary component may overfill their Roche-lobe and form a circumbinary disc \citep{Lubow:2015aa}; we return to this point in the Discussion.

In a standard accretion disc \citep{Pringle:1981aa} around a single object of mass $M$, the matter moves on near-circular orbits with Keplerian orbital frequency $\Omega = (GM/R^3)^{1/2}$ at radius $R$, and the evolution is primarily determined by the action of a viscosity $\nu$, which is typically modelled with the viscous stress proportional to the local pressure \citep{Shakura:1973aa}. Viscous torques acting within the disc are generated by turbulence, which is typically thought to result from the magneto-rotational instability \citep{Balbus:1991aa} and may also occur due to e.g.\@ gravitational instability \citep{Paczynski:1978aa}. These viscous torques transport angular momentum outwards through the disc allowing the mass to spiral inwards, with only a small amount of the mass moving to large radius carrying most of the angular momentum. The magnitude of the viscous torque is given by
\begin{equation}
  \label{nutorque}
  T_\nu = 3\pi\alpha \left(\frac{H}{R}\right)^2\Sigma\Omega^2 R^4\,.
\end{equation}
When matter reaches a radius $R$ it carries specific angular momentum $l = R^2\Omega \approx (GMR)^{1/2}$. Thus, when the matter reaches the disc inner edge (e.g.\@ the surface of the star it is orbiting) and is added to the central object, it gives to that object an amount of angular momentum per unit mass corresponding to the radius from which the matter is accreted. For example, a disc around a black hole has an innermost stable circular orbit, and in general the black hole receives, per unit mass accreted, the angular momentum corresponding to that orbit. If the specific angular momentum of the central object is smaller (larger) than the specific angular momentum of the accreted material then the central object is spun up (down).

For circumbinary discs the most significant additional piece of physics, compared to the standard accretion disc, is orbital resonances (here, specifically, Outer Lindblad resonances), that occur between the disc and binary orbits at discrete locations. Such resonances are the only mechanism by which the binary can remotely transfer angular momentum to the disc \citep{Lynden-Bell:1972aa}. In a circumbinary disc, resonances force the disc orbits to become eccentric with a wave launched outwards, and this leads to a flow of angular momentum and energy to the disc orbits from the binary orbit. These disc motions can be damped by the disc turbulent viscosity, resulting in circular orbits with increased angular momentum, i.e.\@ larger radius \citep{Lin:1979aa} or the waves may be damped locally to the resonance location through non-linear damping effects \citep{Lubow:1998aa}. As the binary has given up angular momentum and energy to the disc, the binary semi-major axis decreases and its orbit may also become (more) eccentric.

Therefore the evolution of the binary is determined sensitively by a competition between these two torques: (1) the capture torque\footnote{We use the word `capture' to imply that the material has been captured from a circumbinary orbit into one around either binary component, thus giving up the circumbinary orbit's angular momentum and energy and adopting approximately the angular momentum and energy of the binary orbit. However, as this material can in principle be subsequently ejected, we do not refer to it as `accreted', but we note that the angular momentum and energy it had in the circumbinary disc has been effectively absorbed by the binary.}, which transfers angular momentum from the material that flows from the circumbinary disc to the binary orbit and (as we will see below) provides an excess of angular momentum per unit mass captured by the binary, and (2) the resonant torque, which transfers energy and angular momentum to the disc from the binary. Each of these torques depends on several parameters of the system and the resulting evolution can therefore be complex.

To understand the specific angular momentum of material captured by a binary from a circumbinary disc, it is important to know the location of the disc inner edge. In their section 3, \cite{Artymowicz:1994aa} give a detailed discussion of the physics responsible for truncation of the inner edge of circumbinary discs. In particular, (their section 3.1) they evaluate the parameters required for the viscous torque to overwhelm the Outer Lindblad resonances \citep[see also][]{Lin:1986aa}. They also discuss alternative nonresonant means of truncating the disc, including a viscous phase lag \citep{Papaloizou:1977aa}, and orbit crossings \citep{Paczynski:1977aa}. Using the methodology of \cite{Papaloizou:1977aa} they estimate the circumbinary disc inner edge location for circular binaries of non-extreme mass ratios, finding that it is at $\approx 1.7a$. This suggests that the location of the inner disc edge is $\gtrsim 1.7a$ for any value of the disc viscosity, and for the cases where the disc viscosity is small enough we can expect the inner disc to be truncated at larger radii by weaker resonances (see equation 16 of \citealt{Artymowicz:1994aa}).

From the above, we can conclude that capture of material from a circumbinary disc, into orbits around one or both of the binary components, leads to the capture of material that has an excess of angular momentum compared to the binary orbit, i.e.\@ the captured material has a higher specific angular momentum than that of the binary. In circumstances where this effect is dominant, it causes the binary to spin up, and therefore to expand with time.

However, as mentioned above, resonances occur between the binary and disc orbits. These transfer both energy and angular momentum from the binary to the disc orbits. The magnitude of the resonant torques can be calculated following \citet[][see e.g.\@ eqns 21-23 of \citealt{Nixon:2015aa}]{Goldreich:1979aa}, and they depend on the binary potential, which is a function of the binary mass ratio and eccentricity, and they are proportional to the disc surface density at the resonance location. Thus, the amount of angular momentum transferred to the disc at each resonance depends on the disc conditions. The angular momentum is transferred to the disc in the form of non-axisymmetric spiral waves, and is predominantly carried by the fundamental mode \citep{Lubow:1998aa}. The surface density of the disc at the location of the resonance is then determined by where the angular momentum carried by the waves is deposited. If the waves can travel outwards by a substantial distance, say of order $R$, then the local surface density is not strongly affected by the resonance and tidal truncation of the disc may be inefficient. However, if the angular momentum is deposited locally in the disc near the location of the resonance, then the disc can be efficiently truncated at the location of the resonance that has the largest radius and is strong enough to impede the accretion flow. It has been shown that in 3D discs with a vertical temperature gradient the waves deposit the angular momentum locally, while in 2D or isothermal discs the waves can travel to large distances \citep{Lubow:1998aa}. Therefore the evolution of the disc-binary system depends sensitively on the system parameters, including the equation of state for the gas, and the viscosity parameter $\alpha$ which can play a role in damping the propagation of waves traveling through the disc.

In general we expect the structure of the circumbinary disc to be thus. Far from the binary we have a smooth disc of material, in which the mass flux may be inwards (`accretion') or outwards (`decretion') depending on the central conditions. At intermediate radii $\gtrsim ka$ where $a$ is the binary semi-major axis and $k$ is of order a few, the disc is somewhat disturbed by outward propagating waves driven by resonant interactions between the disc orbits and the binary. At a radius of $\approx ka$, the disc is tidally truncated and in some (perhaps most) cases this is at the location where circular orbits become unstable and dynamically plunge towards the binary, which occurs at $R\approx 1.7a$ \citep{Artymowicz:1994aa}. Inside the binary orbit, there is a disc of matter around each binary component that is being fed by streams that may form from the circumbinary disc inner edge. The exact location of each of these features, and the surface density profile and masses of the circumprimary/secondary and circumbinary discs, all depend sensitively on the binary-disc parameters.

Following the above discussion, we can make assertions about the disc-binary evolution for different ratios of the viscous to resonant torque strengths. If the viscous torque is sufficiently weak that the resonant torque is able to truncate the disc far from the binary (i.e.\@ at radii $\gtrsim 2a$ where the disc orbits are dynamically stable), then the resulting solution is a decretion disc \citep{Pringle:1991aa}, and the binary must shrink with time. If the viscous torque is sufficiently strong that for any resonance the viscous torque is much stronger than the resonant torque, then we get unimpeded accretion on to the binary orbit (cf. the retrograde case, where the torques are either absent for circular binaries, \citealt{Nixon:2011aa}, or severely weakened for eccentric binaries, \citealt{Nixon:2015aa}), and the disc solution closely resembles the standard accretion disc solution and the binary may, but does not necessarily, expand with time. If the two are comparable (which probably covers several orders of magnitude in the torque ratio) we get time variable accretion and dynamic behaviour within the cavity at $R \lesssim 2a$ and the binary evolution is parameter dependent.

Performing accurate numerical hydrodynamical simulations of a ``true'' decretion disc, where the binary successfully holds the disc out and prevents any accretion, is very difficult, as even for equal-mass circular binaries this requires a sufficiently low viscosity that is difficult to achieve numerically\footnote{We note that it is straightforward in SPH simulations to report the magnitude of the numerical viscosity as one is required to put it in by hand to satisfy the differentiability of the velocity field in the Lagrangian and thus one knows its magnitude. However, in grid codes, and some moving mesh codes, the numerical viscosity is inherent and uncontrolled. For example, the numerical viscosity experienced by an orbiting parcel of gas will depend not only on the resolution, but also on the direction in which the gas parcel traverses the grid. Thus, similarly to the care needed when interpreting SPH simulations of low-density (and thus potentially low resolution) regions of the disc, care must also be taken when interpreting the behaviour of the eccentric inner disc regions in simulations that employ a circular, rather than Cartesian, grid. This point can be seen most sharply in numerical simulations of the tidal instability of circumstellar discs with an external companion that leads to the observed superhump phenomena in e.g.\@ SU UMa type cataclysmic variables \citep[CVs;][]{Warner:1995aa}. Particle based methods recover this dynamics with ease \citep[e.g.\@][]{Whitehurst:1988aa,Lubow:1991aa}, while grid-based methods took substantially longer to reach the same results \citep[e.g.\@][where the results are still subject to numerical details such as the treatment of boundaries]{Kley:2008aa}.}. Some authors have concluded that there is always accretion across the gap, and further that the accretion rate is not slowed compared to the same disc around a single point mass \citep[e.g.\@][]{Shi:2015aa}, but this has not been demonstrated and has been challenged by \cite{Ragusa:2016aa}. However, it is clear that, at least for some parameters, gravitational torques from the binary are insufficient to hold the disc back and accretion on to the binary components occurs \citep[as found by the numerical simulations in][]{Artymowicz:1994aa,Artymowicz:1996aa}, and that both the location of the disc inner edge and the mass flow rate on to the binary are parameter dependent \citep[as found, again, by the numerical simulations in][]{Artymowicz:1994aa,Artymowicz:1996aa}.

In this paper, we are interested in the binary orbital evolution and the recent development that simulations seem to show that the binary expands rather than contracting \citep{Miranda:2017aa,Tang:2017aa,Munoz:2019aa,Moody:2019aa,Munoz:2020aa} contrary to previous works. As has already been pointed out by \cite{Tiede:2020aa}, the simulations which have reported expansion of the binary orbit have been strongly limited in the parameter space they cover. Further they have almost exclusively been performed in 2D (although, we note that \citealt{Moody:2019aa} provide comparison 2D and 3D simulations and find that the flow rate on to the binary differs between them by a factor of five), and we also note that these investigations assume a fixed binary orbit. In the next section we provide two example simulations in 3D with a ``live'' binary that show very different evolution when only one parameter (the disc angular semi-thickness) is varied. We note that the disc-binary evolution is clearly parameter dependent and that subtle changes can affect the evolution significantly, and therefore we will return in the future with a more exhaustive parameter survey.

\section{Numerical simulations}
\label{simulations}
We present three dimensional hydrodynamical simulations of binaries interacting with an external circumbinary disc. We use the publicly available smoothed particle hydrodynamics (SPH) code {\sc phantom} \citep{Price:2018aa}. {\sc phantom} has been used extensively to model disc-binary interactions since \cite{Nixon:2012ac}, see for example \cite{Nixon:2013ab,Facchini:2013aa,Martin:2014aa,Martin:2014ab,Nixon:2015aa,Dogan:2015aa,Martin:2016aa,Kennedy:2019aa}. For the circumbinary discs we simulate here, the gas is modelled with a set of $N_{\rm p}$ particles that are distributed in a disc orbiting the central binary. The disc has an initial inner radius $R_{\rm in}$ and an outer radius $R_{\rm out}$. We take the disc surface density to follow a power-law with $\Sigma = \Sigma_0 (R/R_{\rm in})^{-p}$, where the normalization ($\Sigma_0$) is set by the total disc mass. We model the gas thermodynamics with a locally isothermal equation of state with sound speed $c_{\rm s}=c_{\rm s,0} (R/R_{\rm in})^{-q}$. We model the disc viscosity with a Shakura-Sunyaev $\alpha$ viscosity which is implemented via a direct Navier-Stokes viscosity \citep{Flebbe:1994aa}. We do not include the effects of gas self-gravity \citep[see e.g.\@][]{Cuadra:2009aa}. We model the binary as two equal-mass Newtonian sink particles with accretion radii within which gas particles are removed from the simulation. The binary is initially circular and is ``live'', and therefore responds to both the gravitational torques from the gas disc and conservation of momentum with accreted particles.

For the simulations we present here we have chosen the following parameters. We take the initial disc extent to be from $R_{\rm in} = 3a$ to $R_{\rm out} = 10a$. We employ $R_{\rm in} = 3a$ so that the disc starts close enough to the binary to avoid excessive computational cost following the disc as it moves viscously inwards, but sufficiently far from the binary that we do not start with any material inside the expected tidal truncation radius ($\approx 2a$; \citealt{Artymowicz:1994aa}). We take $R_{\rm out} = 10a$ to allow a disc of sufficient extent that our outer boundary is far from the binary but not so large as to significantly increase computational cost (or at the same number of particles significantly decrease numerical resolution). We do not employ an explicit outer boundary, instead allowing the disc to expand with time. For the surface density and sound speed power-laws we take $p=3/2$ and $q=3/4$, which ensures that the initial disc is uniformly resolved \citep[cf.][]{Lodato:2007aa}. We take the disc mass to be 10 per cent of the binary mass. We take a fixed value for the viscosity of $\alpha=0.3$ (as appropriate for fully ionised discs; \citealt{Martin:2019aa}), and to explore the effects of varying the kinematic viscosity $\nu = \alpha (H/R)^{2} R^{2} \Omega$, and thus varying the torque balance between the binary-driven resonant torque and the disc driven accretion torque, we vary $H/R$ between two simulations as $H/R = 0.03$ (denoted `thin') and $H/R = 0.2$ (denoted `thick'). We note that our choice of parameters leads to $H/R \propto R^{-\delta}$ with $\delta \approx 0.25$, and thus the values of $H/R$ quoted are the values at the initial inner disc radius $R=R_{\rm in}=3a$. Finally we take the accretion radii of the sink particles to be $0.2a$, and thus we expect to accrete most of the particles that orbit inside the binary rather than directly resolving the mini discs that form -- we discuss this further in Section~\ref{discussion}.

\subsection{Results}
\label{results}
We simulate the discs with $N_{\rm p}=10^5$, $10^6$ and $10^7$ particles, finding that in general $10^5$ particles provides marginally resolved simulations, whilst the $10^6$ and $10^7$ particle simulations are adequately resolved for our purposes, with shell-averaged smoothing length per disc scale-height (measured at $R=3a$) of $\left<h\right>/H \approx 0.25~(0.9)$, $0.1~(0.4)$ and $0.05~(0.2)$ for the initial conditions of the thick (thin) discs at $N_{\rm p}=10^5$, $10^6$ and $10^7$ respectively. We have run the simulations with $N_{\rm p}=10^5$ and $10^6$ out to $10,000t_{\rm b}$, where $t_{\rm b}=2\pi(a_0^3/GM_{\rm b})^{1/2}$ is the binary orbital time for the initial binary system. However, we find that on such long timescales ($\gtrsim 1000 t_{\rm b}$) a significant fraction of the disc has been accreted and the discs have expanded to larger radii such that the local resolution near the binary becomes insufficient to adequately resolve the dynamics (the time at which the simulations degrade is earlier for lower numbers of particles). Thus we present our results only up to a time of 1000 binary orbits. For the $10^7$ particle simulations we have been able to run to $\approx 800$ and $\approx 400$ binary orbits for the thin and thick disc simulations respectively. This is sufficient to demonstrate the disc-binary interaction we seek to explore \citep[cf.\@][]{Munoz:2020aa}. As the resolution is determined by the local mass, the resolution changes with radius and time during our simulations. The effect of this can be seen in some of the figures below, with the lowest resolution simulations deviating from the higher resolution runs after a few hundred binary orbits. This typically occurs when the resolution of the $N_{\rm p}=10^5$ run has degraded such that $\left<h\right>/H \gtrsim 1$ as this is where the pressure force becomes overly smoothed and the numerical viscosity dominates over the physical viscosity. For the $10^7$ particle simulations, again measured at $R=3a$, at a time of $t=100t_{\rm b}~(400t_{\rm b})$ we have $\left<h\right>/H \approx 0.15~(0.25)$ and $\approx 0.25~(0.27)$ for the thick and thin cases respectively.

In Fig.~\ref{Fig1} we show the disc structures for the $N_{\rm p}=10^7$ thin and thick simulations (left hand and right hand panels respectively) at several different times through the simulation. The top row shows the initial conditions, and we note that we have used different colour bar ranges for the thick and thin simulations, but have kept the colour bars the same for different times within the same simulation; this is necessary to show the range of features in the surface density at different times. The left hand plots show the thin disc simulation with $H/R=0.03$. After $t=10t_{\rm b}$ (second row), the disc inner edge has moved inwards to $R\approx 2a$, but otherwise remains very similar to the initial conditions. After $t=100t_{\rm b}$ (third row), the disc has reached a quasi-steady state, in which the disc inner edge is strongly affected by tides with outward propagating waves visible in the surface density; significant mass is accumulated at the location of resonances, time-dependent streams of matter feed the binary, and the disc inner edge is both eccentric and precessing. After $t=400t_{\rm b}$ (bottom row, continued overleaf), this quasi-steady behaviour persists. As time passes, the outer disc edge slowly expands.

The right hand plots of Fig.~\ref{Fig1} show the thick disc simulation with $H/R=0.2$, with the initial conditions in the top row. After $t=10t_{\rm b}$ (second row), the thick disc simulation has already reached a settled state. Viewed in the frame corotating with the binary, the disc structure is static and there is no subsequent disc evolution other than a slow decay of the disc surface density with time. The inner disc is circular until a radius that is consistent with the truncation radius predicted by \citet[][see their Table~1]{Artymowicz:1994aa}, at which two spiral arms feed the binary. The disc exhibits no strong signs of resonant interaction with the binary, i.e. no noticeable overdensities are visible at the location of resonances. After $t=100t_{\rm b}$ (third row), and $t=400t_{\rm b}$ (bottom row, continued overleaf), the disc exhibits the same structure but with reduced surface density.

\begin{figure*}[!htp]
  \captionsetup[subfigure]{labelformat=empty}
  \subfloat[][]{\includegraphics[width=0.498\textwidth]{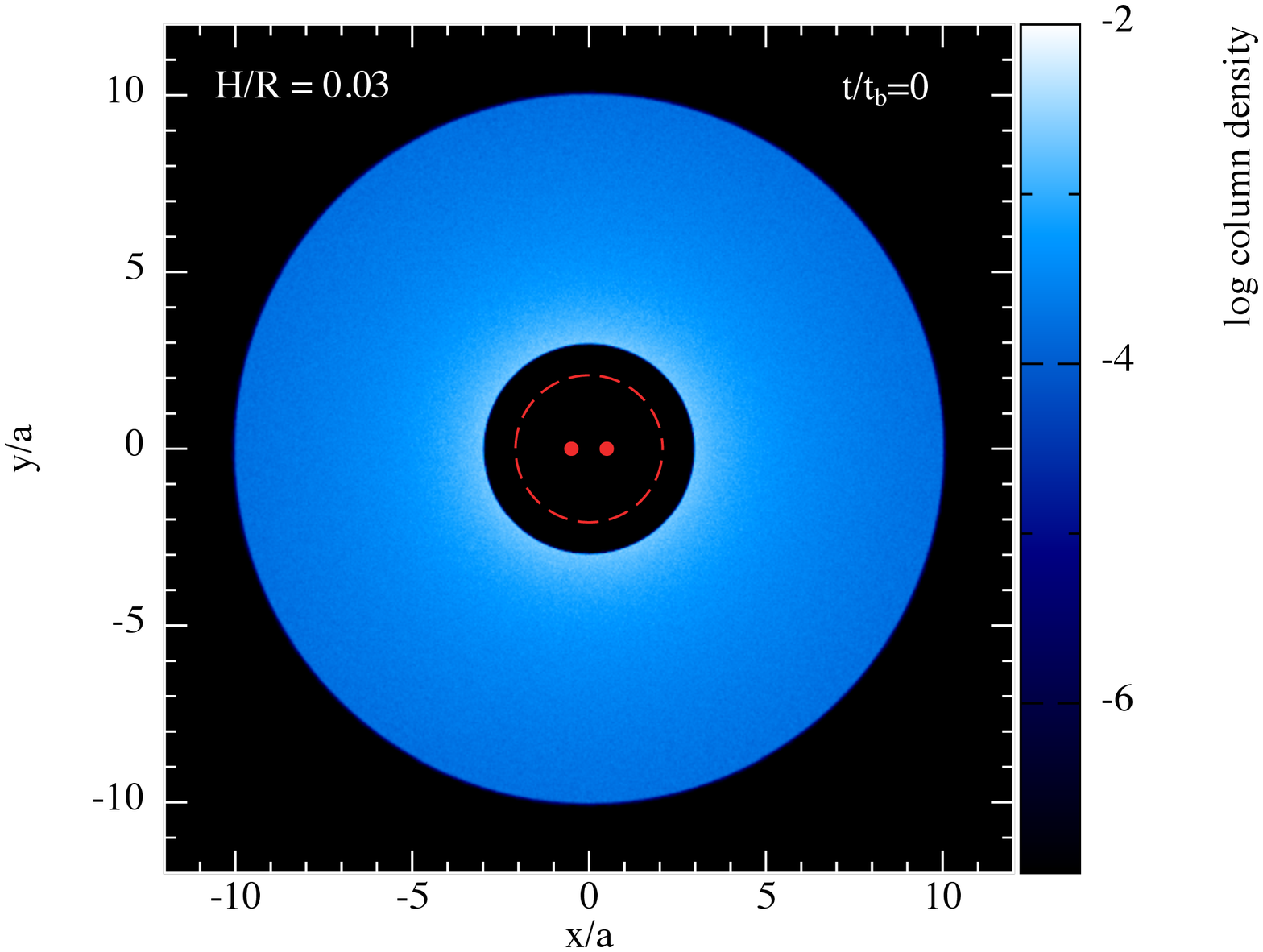}}\hfill
  \subfloat[][]{\includegraphics[width=0.498\textwidth]{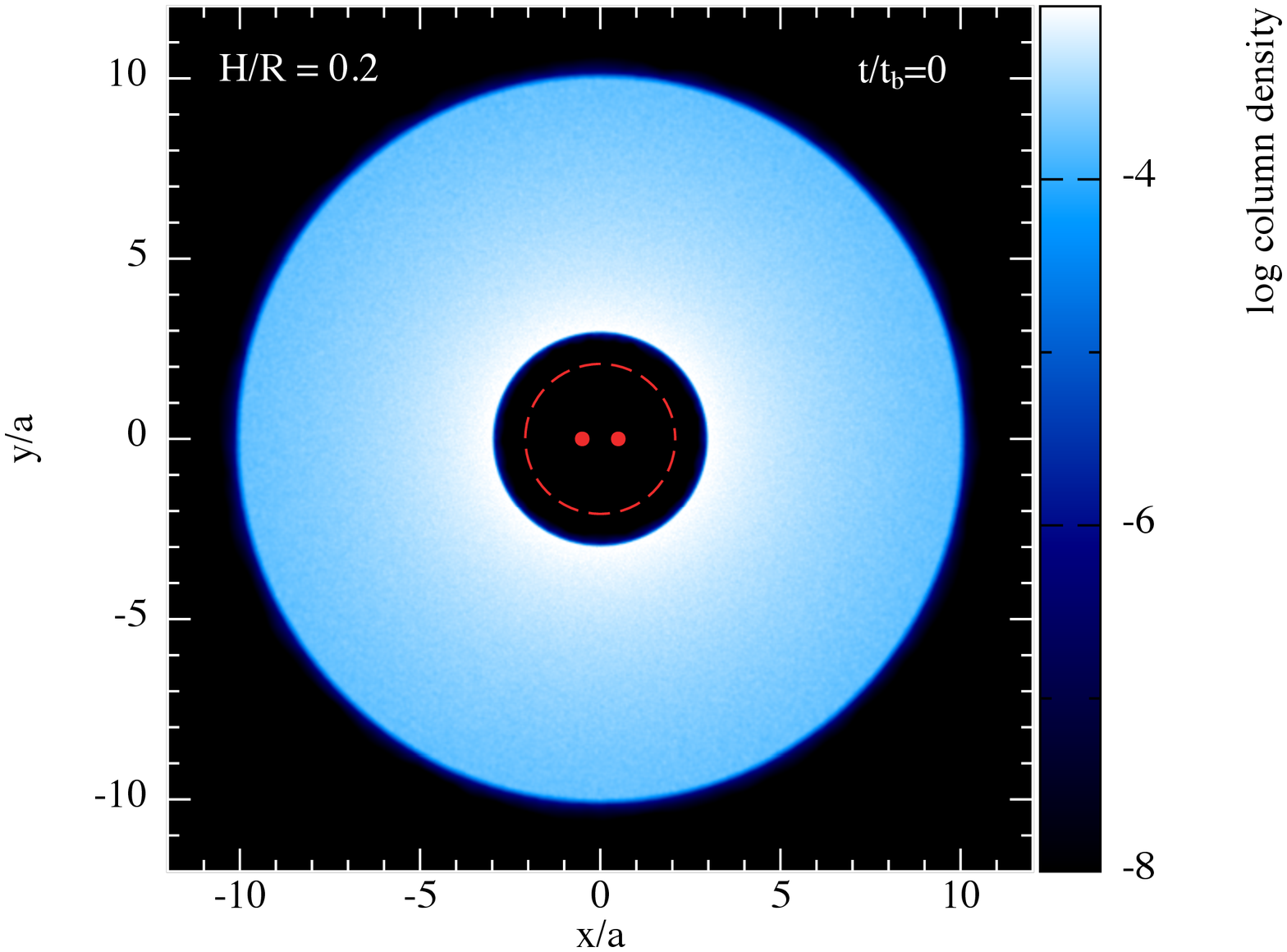}}\\
  \subfloat[][]{\includegraphics[width=0.498\textwidth]{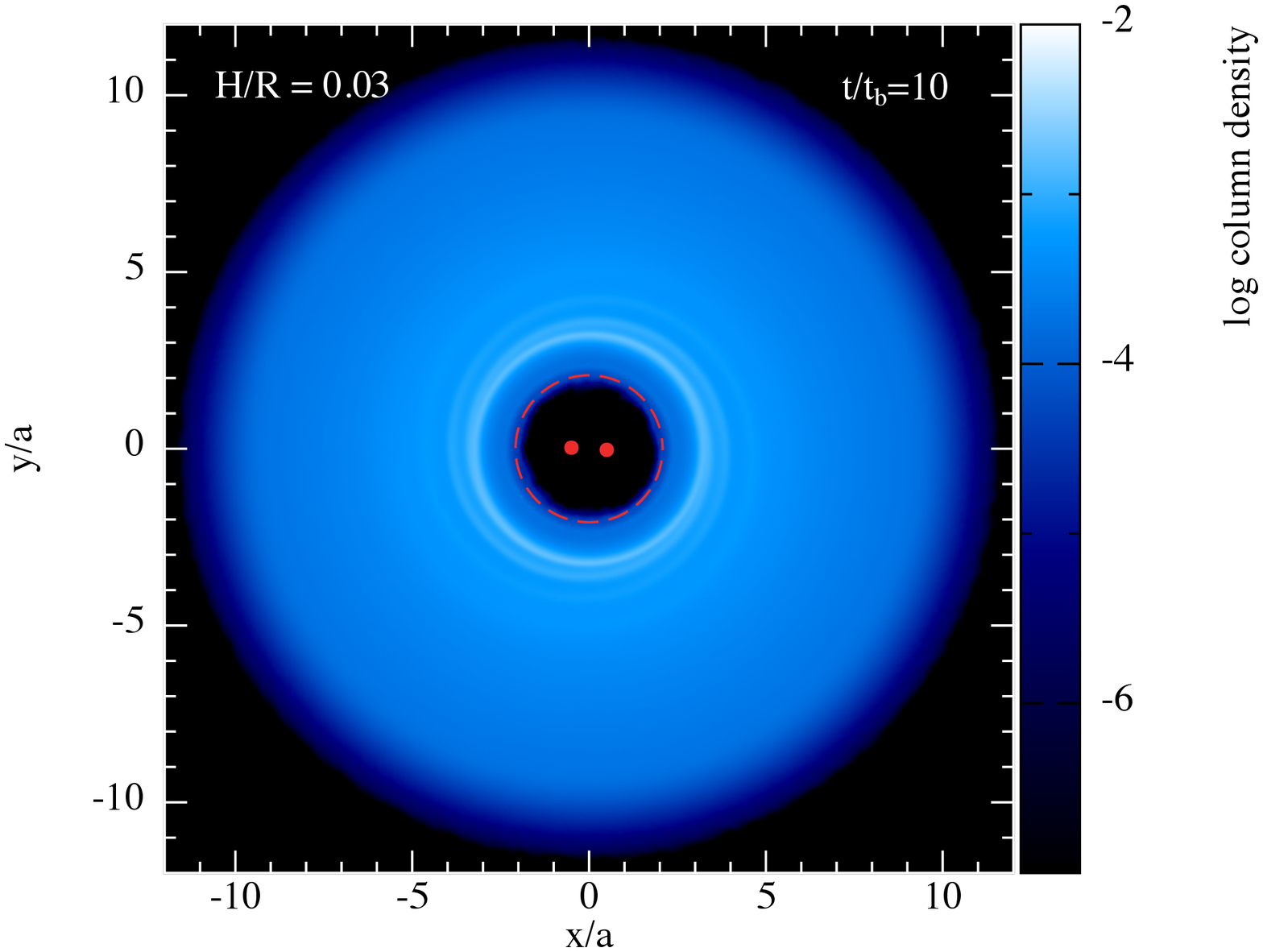}}\hfill
  \subfloat[][]{\includegraphics[width=0.498\textwidth]{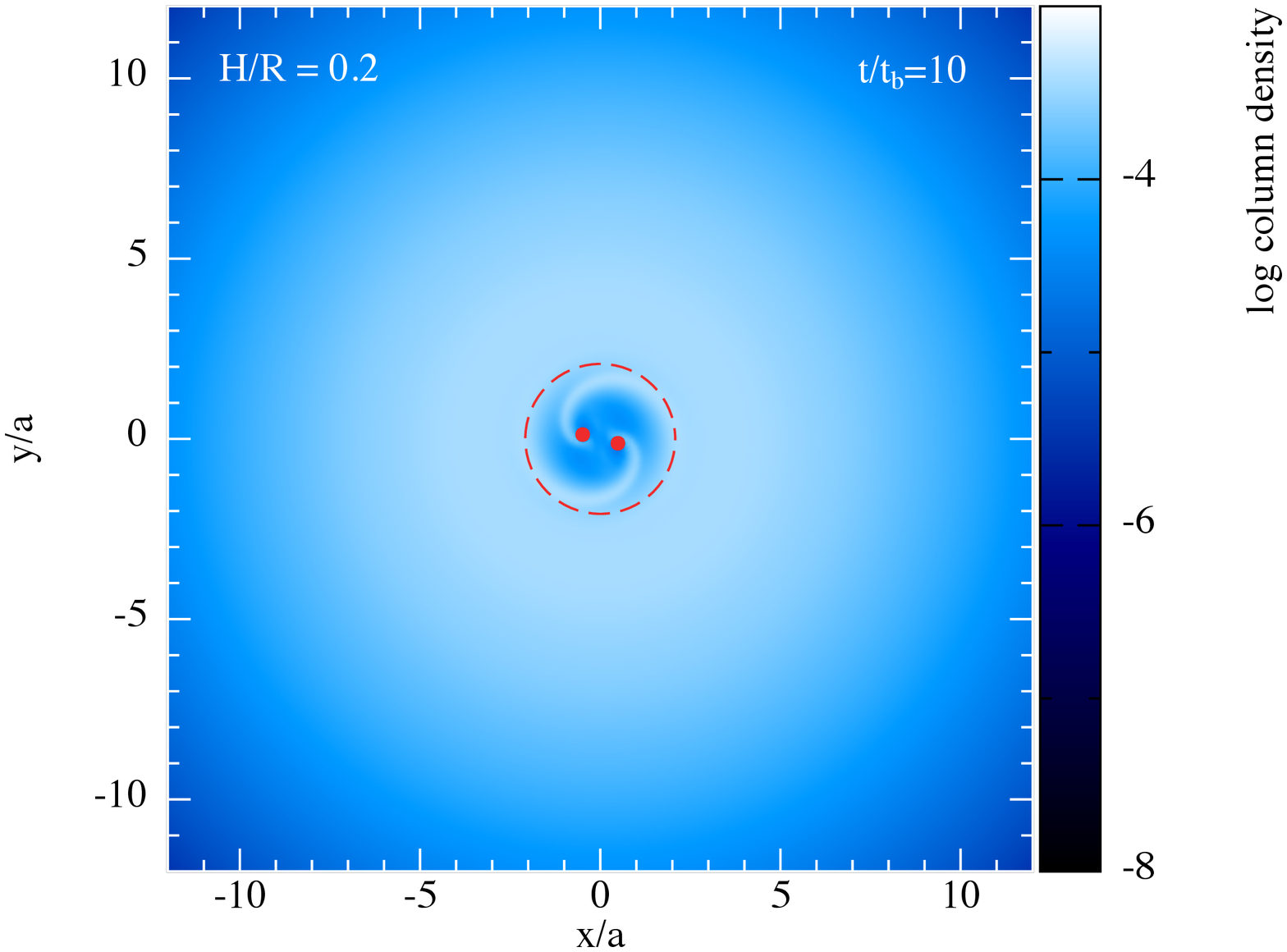}}\\
  \subfloat[][]{\includegraphics[width=0.498\textwidth]{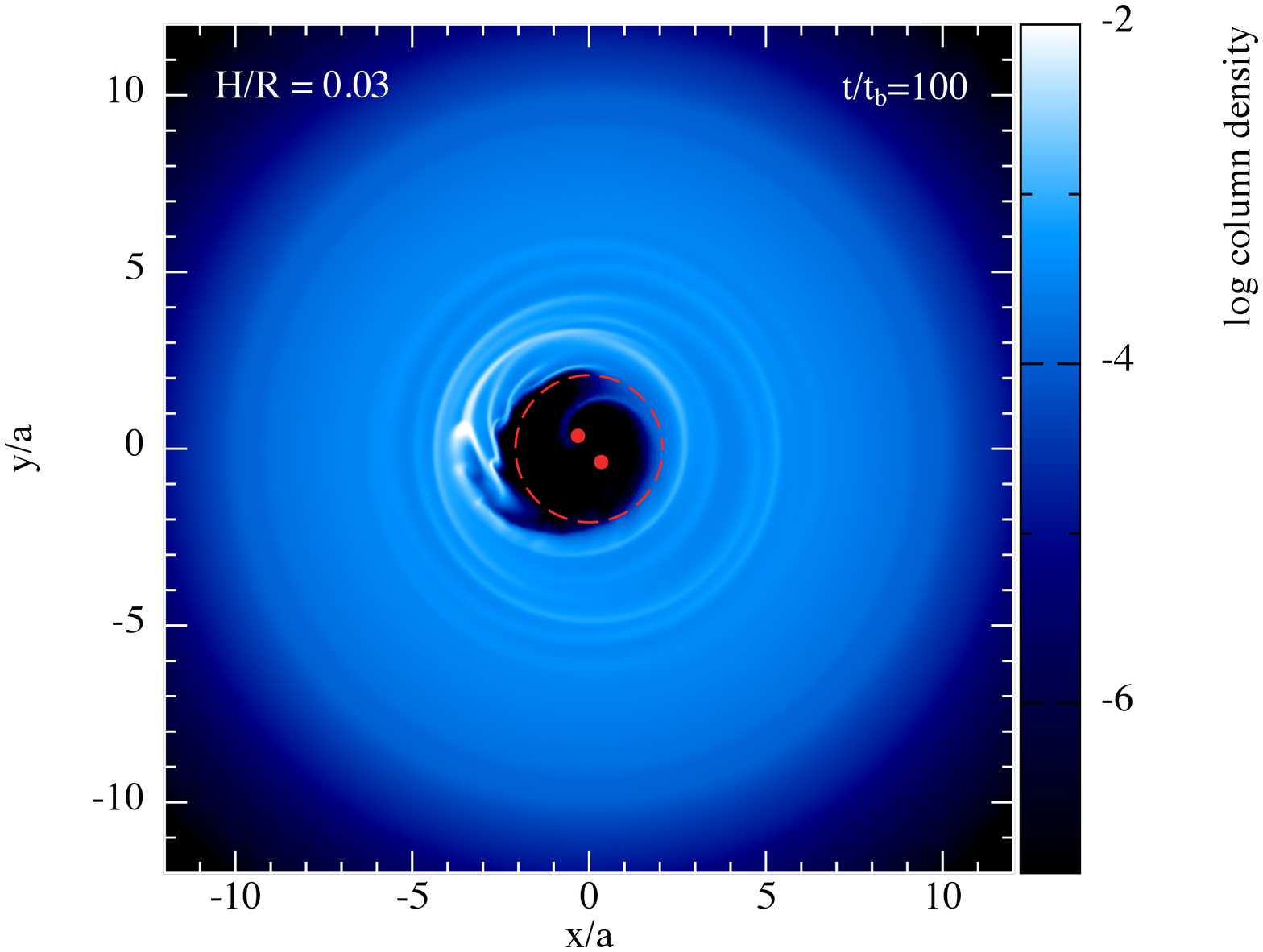}}\hfill
  \subfloat[][]{\includegraphics[width=0.498\textwidth]{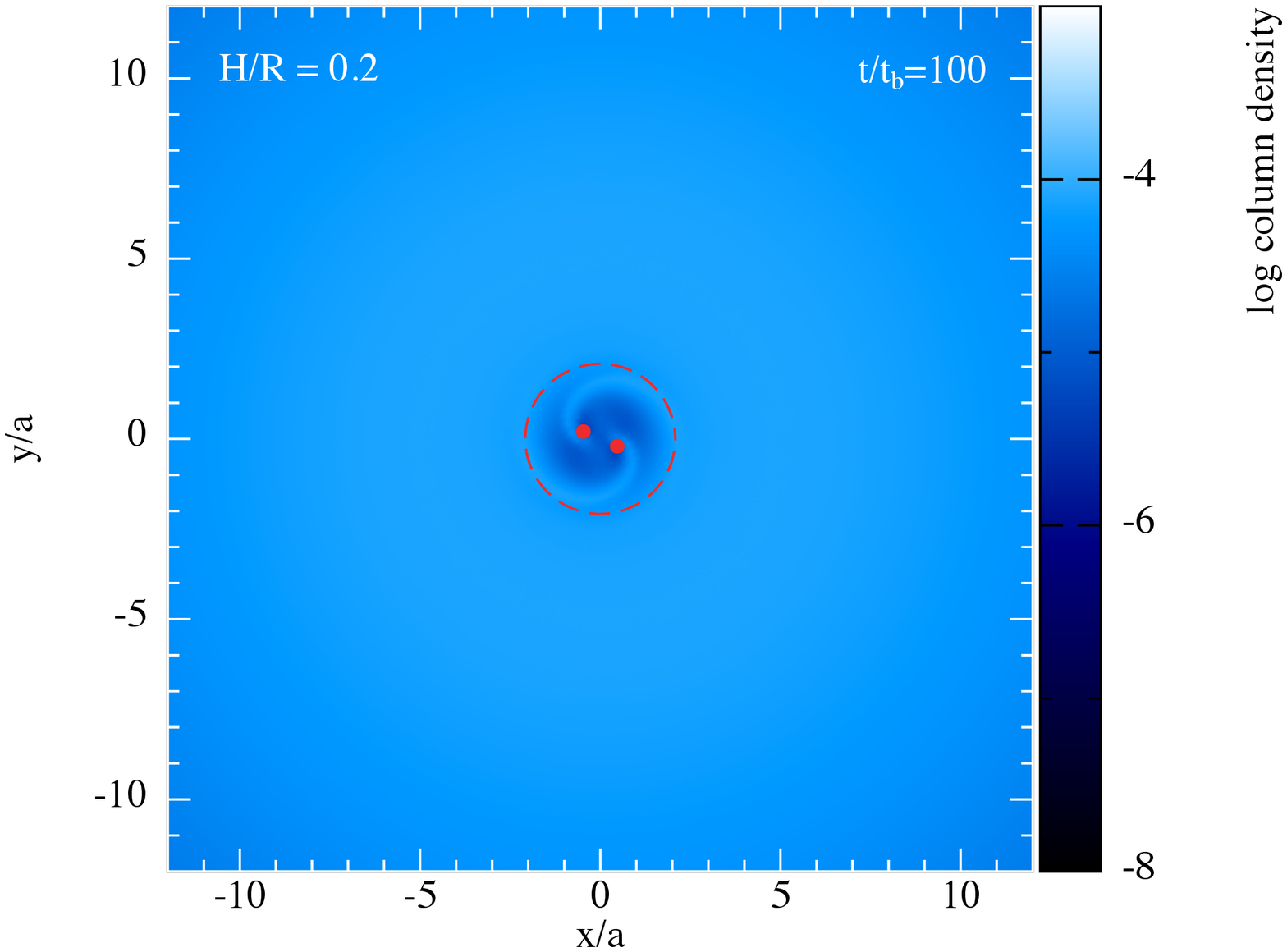}}\vspace{-0.3in}
  \caption{Column density plots for the thin ($H/R = 0.03$; left hand panels) and thick ($H/R = 0.2$; right hand panels) disc simulations. (\emph{Figure and caption continued overleaf.})}
  \label{Fig1}
\end{figure*}
\begin{figure*}[!htp]
  \captionsetup[subfigure]{labelformat=empty}
  \ContinuedFloat
  \subfloat[][]{\includegraphics[width=0.498\textwidth]{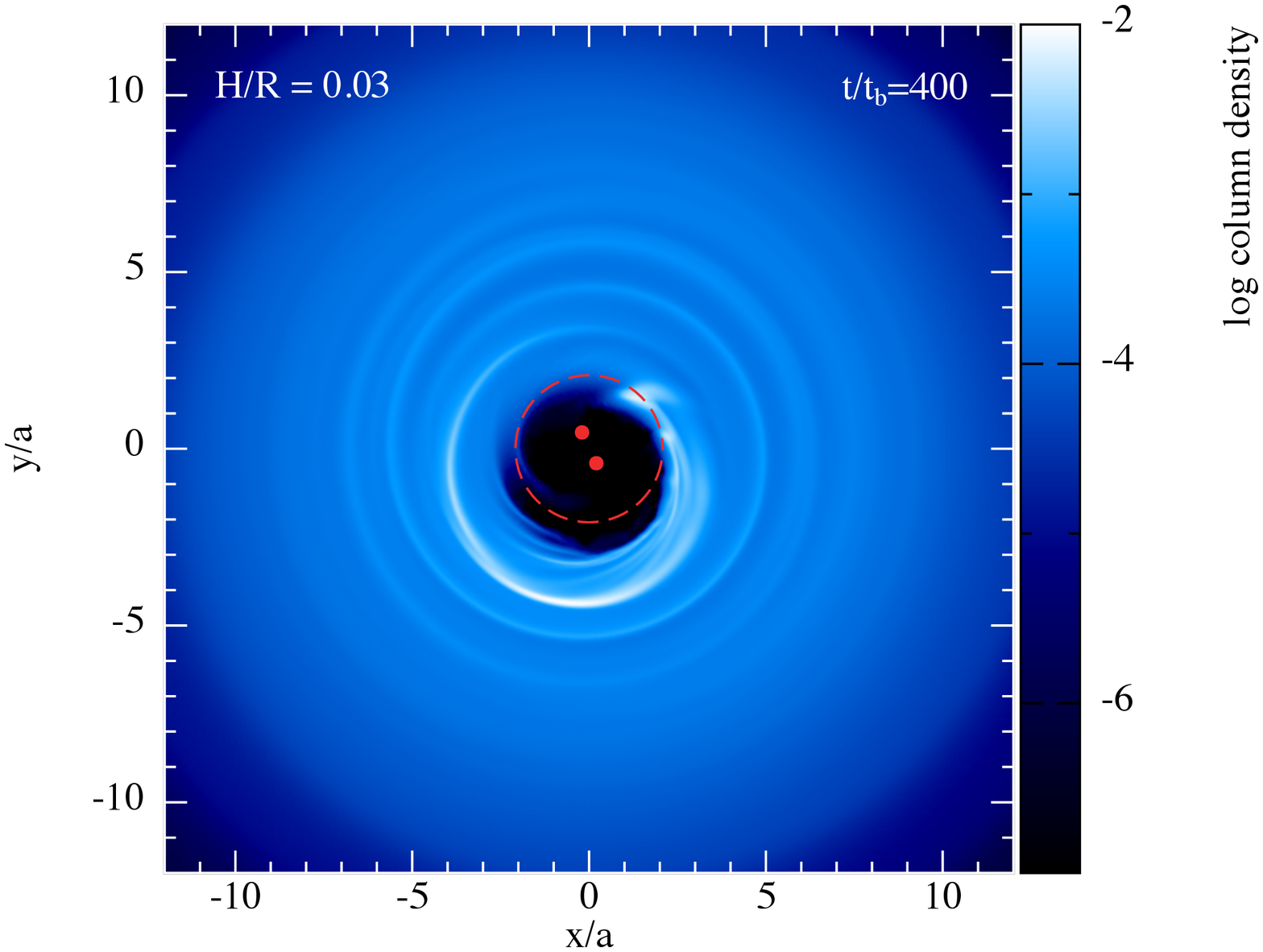}}\hfill
  \subfloat[][]{\includegraphics[width=0.498\textwidth]{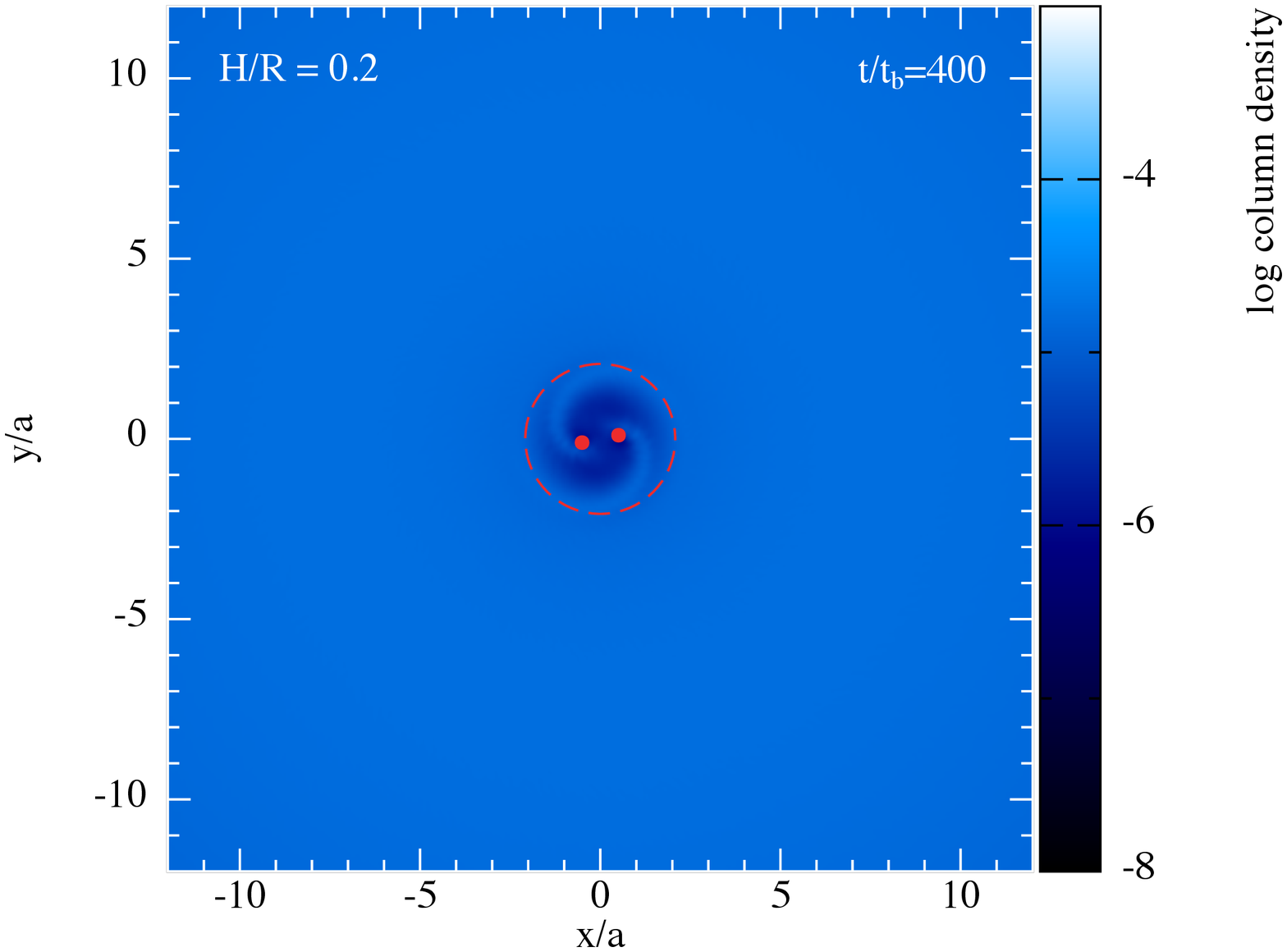}}\vspace{-0.3in}
  \caption{{\it Continued from previous page.} Column density plots for the thin ($H/R = 0.03$; left hand panels) and thick ($H/R = 0.2$; right hand panels) disc simulations. These images are made at times of $0t_{\rm b}$ (top row), $10t_{\rm b}$ (second row), $100t_{\rm b}$ (third row) and $400t_{\rm b}$ (4th row), where $t_{\rm b} = 2\pi (a_0^3/GM_{\rm b})^{1/2}$ is the initial orbital time of the binary. Note that as the binary is allowed to evolve in the simulations, the binary has actually executed a slightly larger (smaller) number of orbits in the thin (thick) disc case. The colour bar denotes the column density, and thus the view is of the $x$-$y$ plane (i.e.\@ the binary orbital plane) and the density has been integrated through all $z$. Note that the colour bars are the same for all of the thin disc panels and for all of the thick disc panels, but are different between the two simulations; this is because the two simulations show different evolution of the surface density, with the thin disc case producing increased surface density at the locations of strong disc resonances (the white regions at $t=100t_{\rm b}$ and $t=400t_{\rm b}$) whereas the thick disc case shows a rapid decline in disc surface density due to significant accretion on to the binary and the disc expanding to larger radii. The disc morphology is distinct between the thick and thin disc cases. The thick disc quickly establishes a steady-state morphology comprising a smooth (accretion) disc for $R\gtrsim 2a$ with two strong streams of material spiraling on to the binary at $R\lesssim 2a$. As time proceeds this morphology persists and the surface density decays. For the thin disc a quasi-steady morphology takes longer to occur, with, on timescales $t\gtrsim 100 t_{\rm b}$, a quasi-steady, eccentric, and precessing inner disc edge at $R\approx 2-3 a$ is observed. This structure persists for the remainder of the simulation (to $\approx 1000 t_{\rm b}$). For circumbinary discs \citep[see][]{Artymowicz:1994aa} the strongest resonance that is generally expected to be in the disc is given by the $(m,l) = (2,1)$ resonance (where $m$ is the azimuthal number and $l$ is the time-harmonic number), which occurs at a radius of $2.08a$; this location is marked on each panel with a red dashed circle.} 
 \end{figure*}

\begin{figure*}[!htp]
  \includegraphics[width=0.485\textwidth]{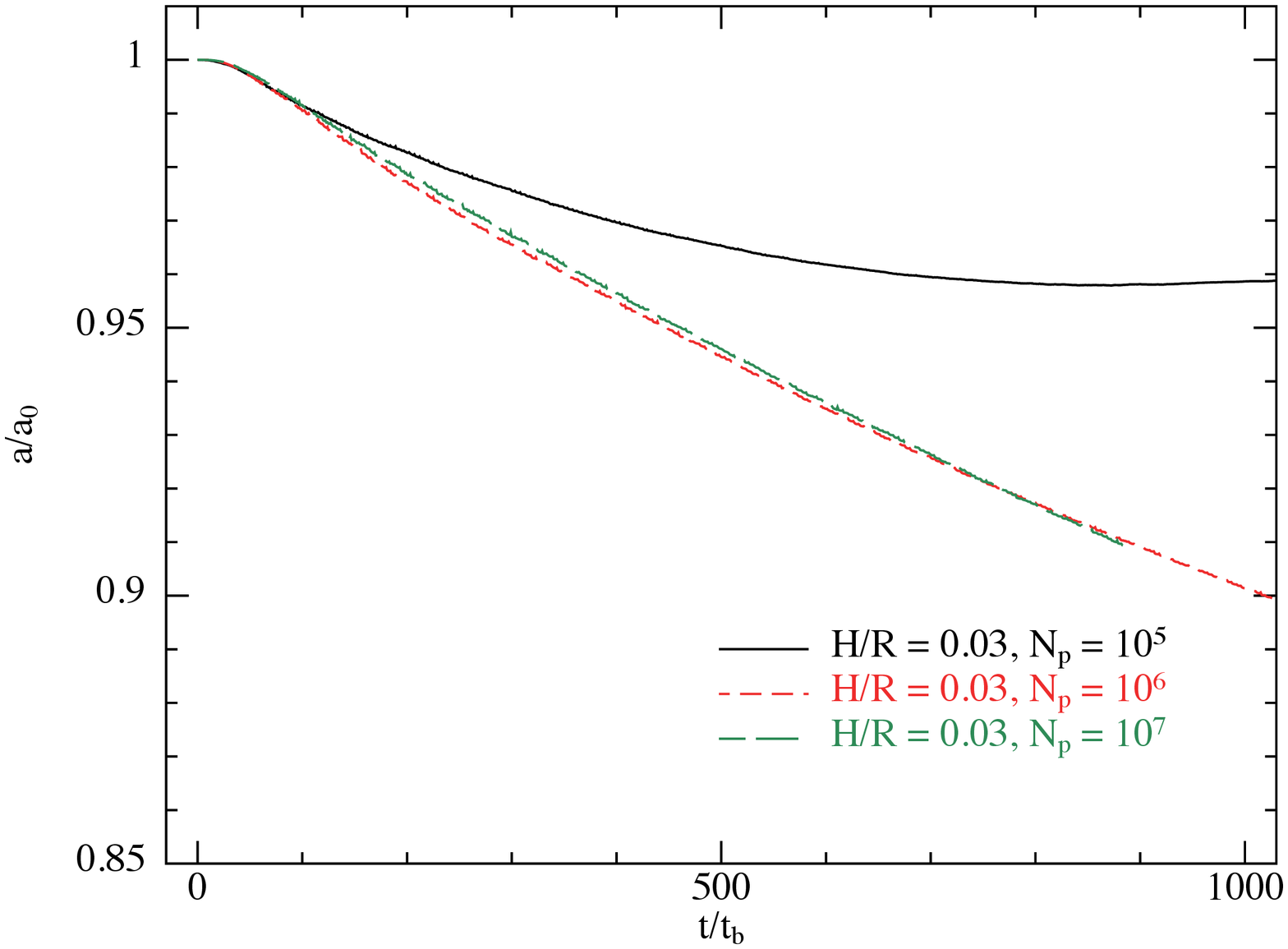}\hfill
  \includegraphics[width=0.485\textwidth]{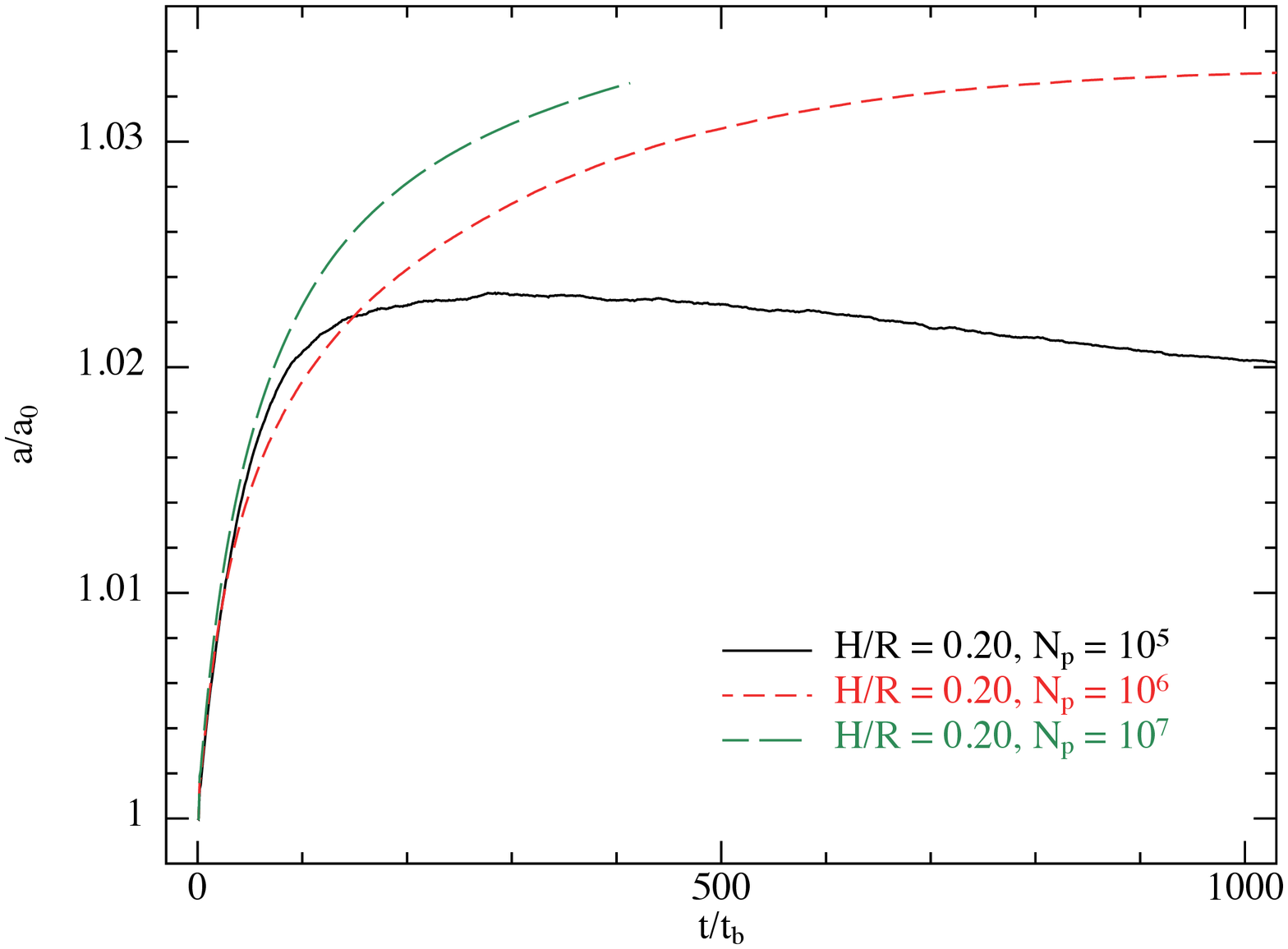}
  \includegraphics[width=0.485\textwidth]{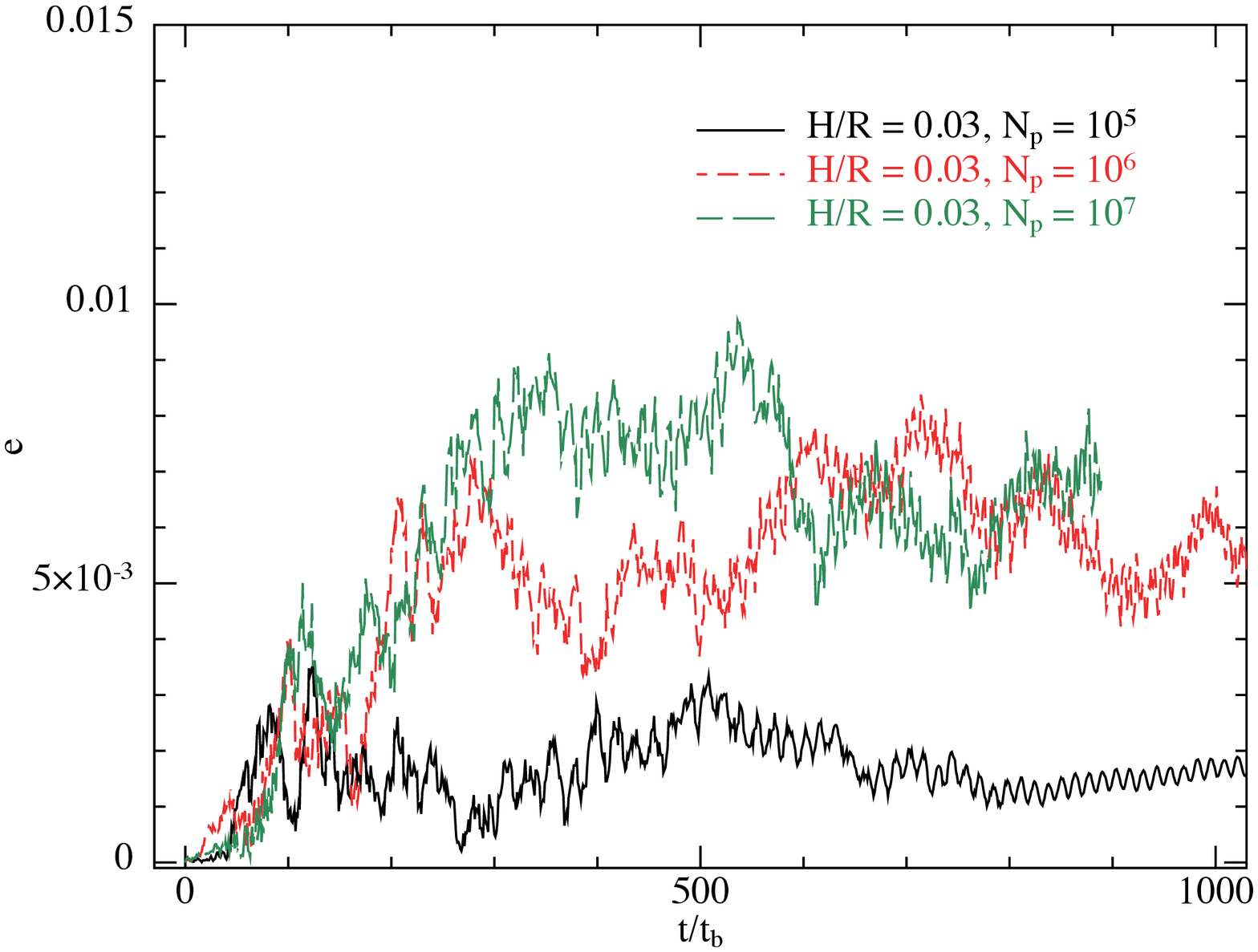}\hfill
  \includegraphics[width=0.485\textwidth]{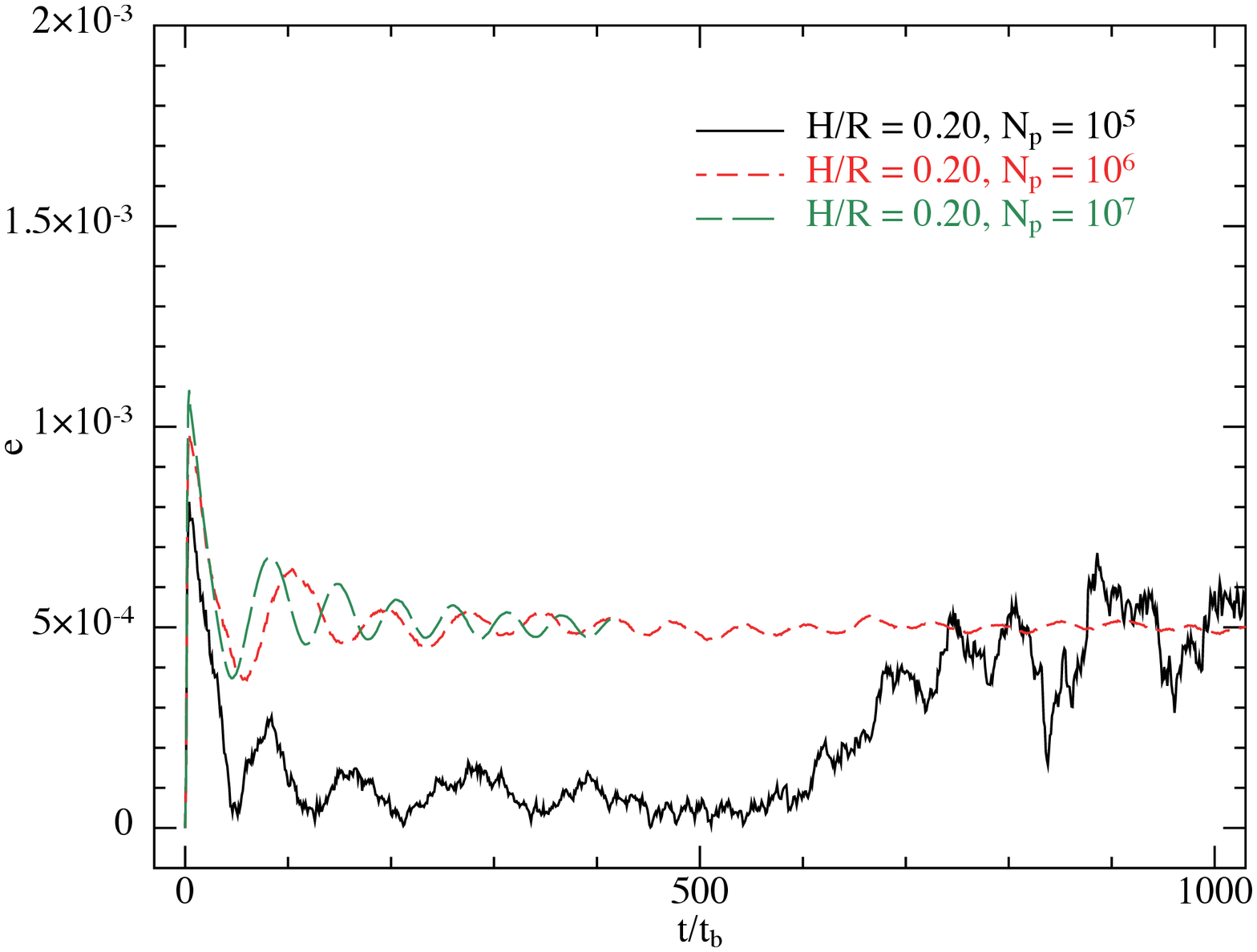}
  \caption{Time evolution of the binary semi-major axis (top panels) and eccentricity (bottom panels) for the thin disc (left hand panels) and thick disc (right hand panels) cases, at three different resolutions (black solid line $N_{\rm p}=10^5$, red dashed line $N_{\rm p}=10^6$, and green long dashed line with $N_{\rm p}=10^7$).  For the thin disc case, at each resolution the binary shrinks. For higher resolution the behaviour of the binary orbit appears to be converged.  While at lower resolution, the binary orbit starts to expand at late times due to the resolution length scale increasing with time, which correspondingly leads to a large numerical viscosity at late times. For the thick disc case the binary expands with time and the amount of expansion increases with increasing resolution. While the higher resolution cases show the same behaviour, after several hundred binary orbits the lowest resolution case exhibits binary orbital decay. We attribute this to a lack of resolution at these late times which correspondingly moves the inner edge of the circumbinary disc to larger radii which in turn weakens the capture torque and allows tides to become artificially dominant in this case. In both the thick and thin disc cases the binary eccentricity remains small ($e\lesssim 0.01$), but the thin disc case, which shows a larger asymmetry in the gas distribution, has a significantly higher eccentricity than the more symmetric thick disc case.}
  \label{Fig2}
\end{figure*}

Fig.~\ref{Fig2} shows the evolution of the binary orbit with time. As we evolve the binary orbit in our simulations, we can directly plot the binary semi-major axis and eccentricity with time. The top left panel shows the evolution of the semi-major axis for the thin disc simulations at all three resolutions. Initially, for $t\lesssim 200t_{\rm b}$, all three resolutions provide the same evolution; the binary shrinks with time. For $t \gtrsim 100t_{\rm b}$ the lowest resolution simulation $N_{\rm p}=10^5$ has inadequate resolution with the binary orbit stalling for $500 \lesssim t/t_{\rm b} \lesssim 1000$ and expanding very slowly after this. In contrast, the higher resolution simulations show very similar behaviour to each other and the binary continues to shrink for the full duration of both simulations. The bottom left panel of Fig.~\ref{Fig2} shows the time evolution of the eccentricity in this case. For all three resolutions the eccentricity remains small ($e\lesssim 0.01$) for the duration of the simulations. The top right panel of Fig.~\ref{Fig2} shows the evolution of the semi-major axis for the thick disc simulation at each resolution. Again, each simulation shows the same evolution for $t\lesssim 100t_{\rm b}$; the binary expands. At later times, the lowest resolution simulation becomes poorly resolved and the binary expansion stalls and later begins to contract. In contrast, the highest resolution simulations continue to show the binary expanding and this continues throughout the duration of the simulations. The bottom right hand panel of Fig.~\ref{Fig2} shows the eccentricity evolution for the thick disc case. Here the binary remains circular throughout ($e\approx 5\times 10^{-4}$).

\begin{figure*}[!htp]
  \includegraphics[width=0.485\textwidth]{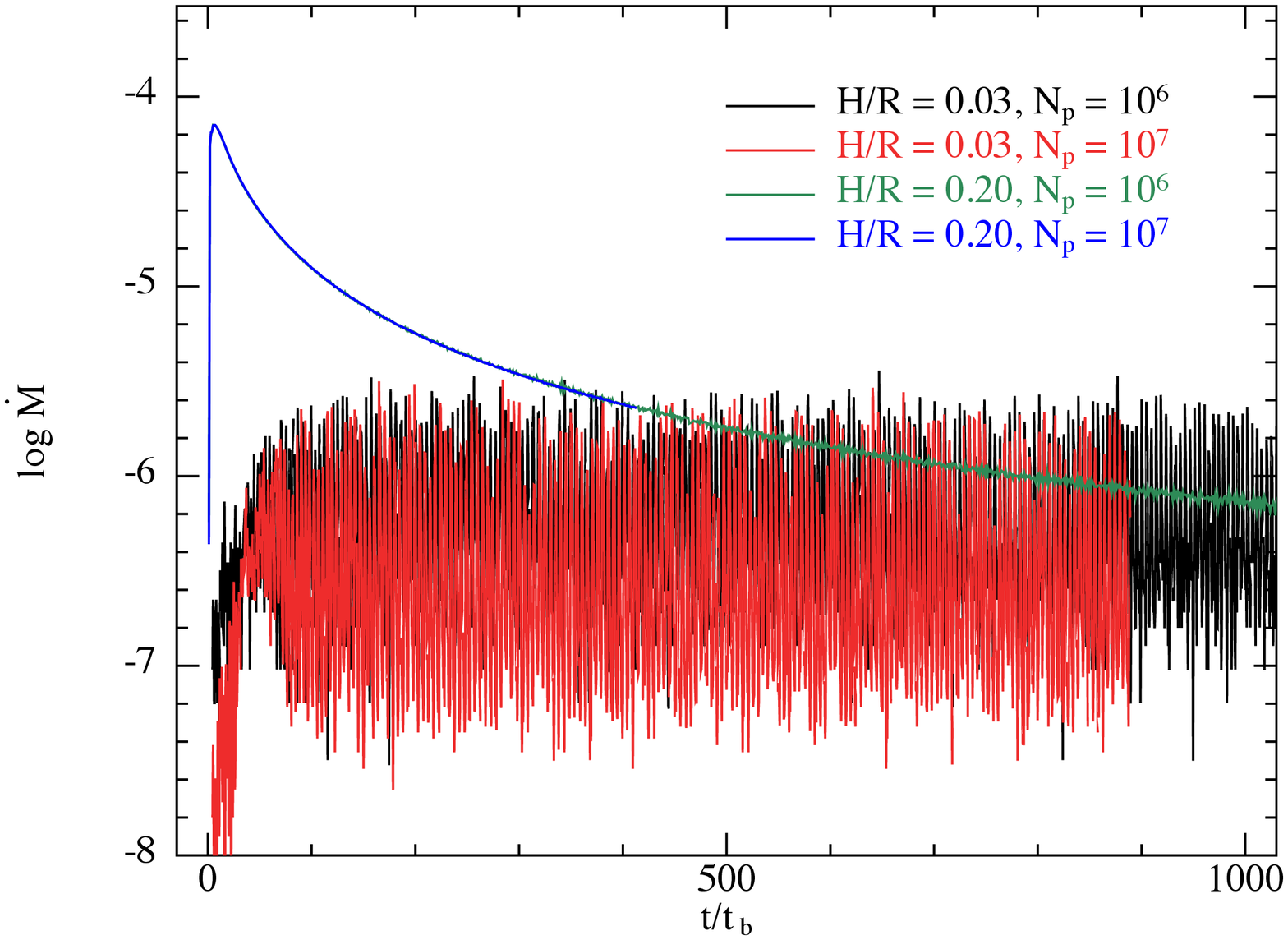}\hfill
  \includegraphics[width=0.485\textwidth]{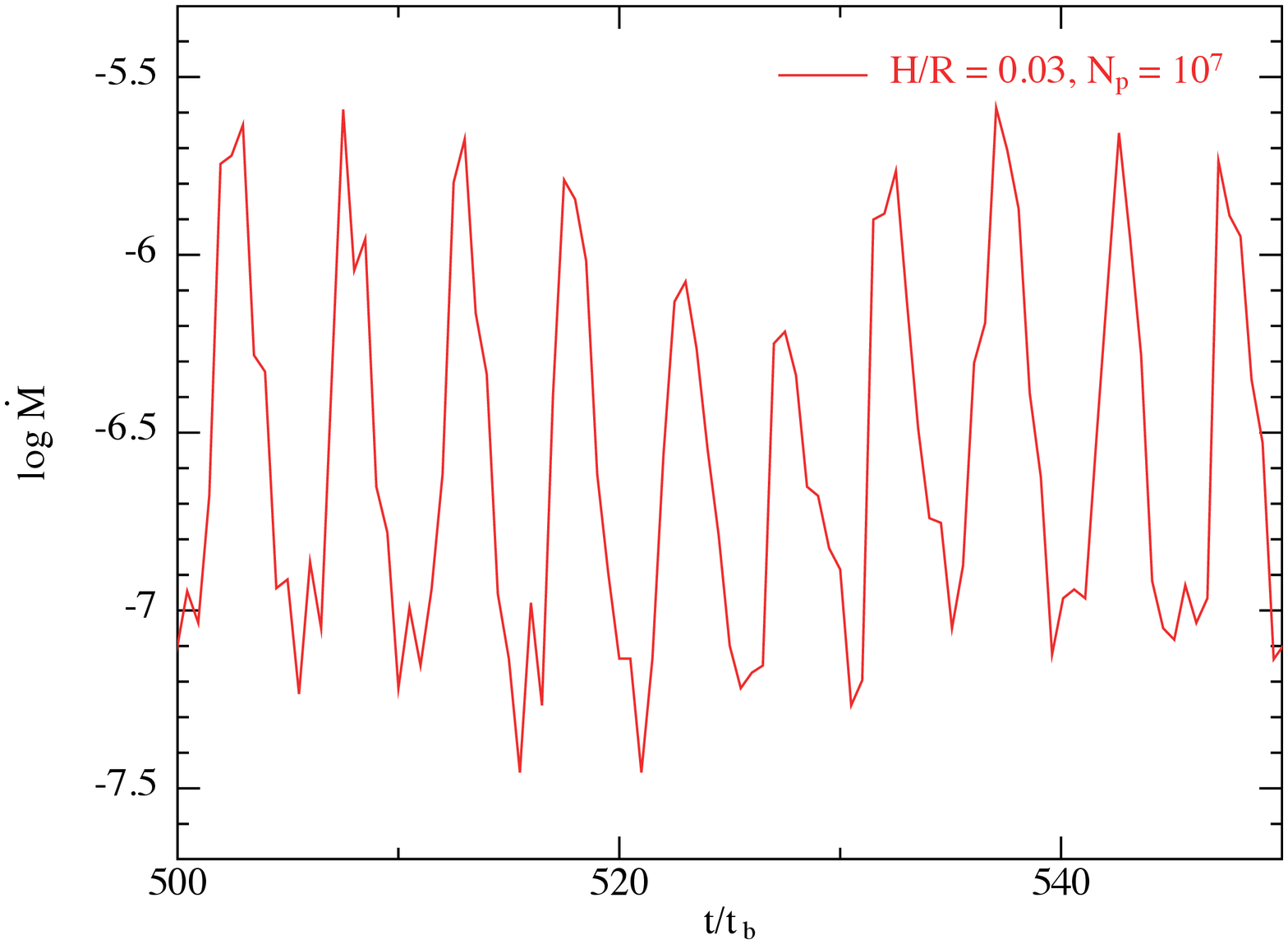}
  \caption{Time evolution of the accretion rate on to one of the components of the binary (recall that the binary is equal mass and near-circular and therefore the accretion rate on to each binary component is very similar). The left hand panel compares the accretion rates for the thick and thin disc cases and at two different resolutions. The black line corresponds to the simulation with $H/R=0.03$, $N_{\rm p}=10^6$, the red line corresponds to the simulation with $H/R=0.03$, $N_{\rm p}=10^7$, the green line corresponds to the simulation with $H/R=0.2$, $N_{\rm p}=10^6$, and the blue line corresponds to the simulation with $H/R=0.2$, $N_{\rm p}=10^7$. For the thick disc case (blue \& green lines) the accretion rate declines with time monotonically following the usual viscous decay of the disc surface density (cf. the right hand panels of Fig~\ref{Fig1}). We note that the blue line ($N_{\rm p}=10^7$) lies directly on top of the green line ($N_{\rm p}=10^6$) demonstrating that the accretion rate is converged in this case. For the thin disc case the accretion rates show the same morphology between the two resolutions, but with a significantly different shape from the thick disc case. The binary accretion rate for the thin disc is highly variable, varying by up to 2 orders of magnitude on a timescale of $\approx 5t_{\rm b}$. The right hand panel shows a zoom in of the accretion rate on to the binary in the thin disc case for $N_{\rm p}=10^7$ between 500 and 550 binary orbits. The periodic nature can be seen clearly along with a secular variation corresponding to precession of the disc inner edge \citep[see also e.g.\@][]{Macfadyen:2008aa}.}
  \label{Fig3}
\end{figure*}

Fig.~\ref{Fig3} shows the accretion rates---that is the mass flow rate through the sink radius---on to one of the binary components with time. For the thick disc case, this shows a very simple curve that peaks at the start and falls off with time as the disc surface density decays due to both accretion and spreading of the disc to larger radii. For the thin disc case, reflecting the dynamical behaviour of this disc near the binary, we find a strongly variable mass flow rate on to the binary that, for each binary component, varies by up to two orders of magnitude on timescales $\approx 5t_{\rm b}$, which corresponds to the precession timescale of the eccentric inner disc. A zoom-in of the accretion rate in the thin disc case is shown in the right hand panel. Fig.~\ref{Fig3} shows that the peak accretion rate for the thick disc is several orders of magnitude higher than that of the thin disc case, with the difference being a factor of a few higher than implied by the ratio of the viscosities in each case \citep[consistent with e.g.\@][]{Ragusa:2016aa}. Therefore, we find that the thick disc case closely resembles an `accretion' disc as the gravitational torque applied to the disc is small compared to the accretion torque. While for the thin disc case the torque applied to the disc inner region is strong resulting in only modest levels of accretion and matter being viscously expelled through the disc to larger radii. Thus, in terms of the ``$f$'' parameter proposed by \cite{Nixon:2020ab} to describe discs with a non-zero central torque, we find that the thick case has $f \ll 1$ and the thin disc case has $f \gg 1$.

\section{Discussion}
\label{discussion}
We have highlighted that the dynamics of circumbinary discs is complex and subtle, depending on several parameters whose effects interplay to give the resulting physical behaviour. In this section we provide discussion on various aspects of the problem. In the next subsection we discuss the role that each system parameter plays in determining the disc-binary evolution.

\subsection{Dependencies on physical parameters}
The total mass of the binary and the binary semi-major axis both determine the physical timescale of the system, i.e.\@ the binary orbital time, and thus (assuming that the equation of state has no explicit time dependence) the problem is scalable in these quantities. However, the binary eccentricity and mass ratio cannot be scaled out of the problem as they determine the location and strength of resonances \citep{Artymowicz:1991aa}. Therefore these parameters play a role in determining the location of the disc inner disc, through balancing the disc viscous torque or determining where orbits become unstable and plunge on to the binary. Thus these parameters affect the efficiency of accretion and the possibility of binary expansion, and also the time dependence and distribution of the material captured by the binary.

The properties of the disc primarily enter through the viscous torque (equation~\ref{nutorque}). Here the key parameters are $\alpha$ and $H/R$, with the combination $\alpha(H/R)^2$ determining the magnitude of the viscous torque in the disc, and thus affecting the mass flow rate on to the binary and the location of the disc inner edge. \cite{Artymowicz:1994aa} argue that for thin discs the relevant parameter is the torque coefficient $\alpha(H/R)^2$, but they also note that for sufficiently thick discs the wavelength of density waves is increased, resulting in a less localised torque and a smoother density profile\footnote{We suggest here, that this effect is the reason why \cite{Tiede:2020aa} find that the binary orbit can expand or contract when $\alpha(H/R)^2$ is kept constant, but $H/R$ is varied. For the thicker disc cases, which have a larger pressure lengthscale, the resonant torques are less able to extract angular momentum from the binary orbit as mass is less strongly concentrated at the location of the resonances.}. It therefore seems likely that $\alpha$ and $H/R$ become distinct parameters in determining the evolution for the thick discs in which binary expansion has been found. Additional disc parameters include quantities concerning the equation of state, e.g. the gas cooling timescale or the adiabatic exponent. These affect the propagation of waves in the disc and thus the location at which the energy and angular momentum, that is communicated to the disc through resonances, is deposited \citep{Lubow:1993aa,Korycansky:1995aa,Lubow:1998aa,Bate:2002aa}\footnote{We note that the propagation of waves, and thus the disc response is markedly different between 2D and 3D discs, and between isothermal and polytropic discs. Thus care should be taken interpreting the behaviour of physical systems from numerical simulations that employ such approximations.}. Similarly, the value of $\alpha$, which damps propagating waves, has an affect here, but perhaps only at high values of $\alpha$ as the waves are predicted to dissipate their energy within a few $H$ of the resonance location for realistic discs \citep[e.g.\@][]{Lubow:1998aa,Bate:2002aa}. Additionally, the disc mass may play a non-trivial role in determining the secular disc evolution; primarily the disc mass determines the timescale on which the binary orbit evolves, but it is possible that non-linear coupling between the binary orbital evolution and the disc means that a simple rescaling of solutions for different disc masses is not possible. Finally, we have not discussed here the possibility that the disc is misaligned or retrograde \citep[we refer the reader to e.g.\@][for these cases]{Larwood:1997aa,Ivanov:1999aa,Nixon:2011aa,Nixon:2013ab,Ivanov:2015aa,Nixon:2015aa}.

\subsection{Numerical considerations}
Numerically this is a very difficult problem, and in many respects it is similar to planet migration through protostellar discs for which the field has not yet converged on a clear answer \citep{Dangelo:2005aa,Paardekooper:2006aa,Nelson:2018aa,Armitage:2019aa}. To faithfully represent disc-binary interaction with numerical hydrodynamics one requires sufficient spatial scale to capture resonant locations and a sufficiently large outer radius that any waves which are launched do not interact with the outer boundary and return to the binary. One requires sufficient resolution to capture the excitation of waves and their propagation, and sufficient run times for the system to relax to a (quasi-)steady state. One may also require sufficient resolution to resolve the accretion discs that form around the binary components, although we consider this a less important requirement as once the circumbinary disc material is captured by the binary components, the angular momentum and energy ultimately ends up with the binary orbit through accretion or tides (unless material from each disc can return to the circumbinary disc which would serve only to, at most, extract additional energy and angular momentum from the binary orbit). Further, one requires careful control of numerical viscosity (either explicit in Lagrangian codes or implicit in Eulerian codes), particularly for non-circular orbits near the binary. And then, as discussed above, there are several independent parameters to vary, and more if one doesn't assume a simple equation of state.

Perhaps the most important physical question, that must be disentangled from numerical effects, is: where in the disc is the energy and angular momentum, that is extracted from the binary through resonances, deposited? This is because the location and rate at which the transferred energy is dissipated, and the angular momentum deposited, directly governs the disc response to the imposed torque and thus controls the solution to the problem. Recall that the location differs when modelling the disc in 2D or 3D, and differs between isothermal equations of state and e.g.\@ polytropic \citep{Lubow:1993aa,Korycansky:1995aa,Lubow:1998aa,Bate:2002aa}, with the often more realistic case (3D and not isothermal) showing that the energy and angular momentum is deposited close to the resonance location \citep{Lubow:1998aa,Bate:2002aa}. If the energy and angular momentum is deposited near the binary, then the inner disc regions are strongly affected, and efficient truncation seems likely. However, if the dissipation occurs only far from binary (as occurs in 2D or isothermal discs), then truncation is less efficient and significant amounts of matter may be allowed to reach the binary.

Recently \cite{Munoz:2020aa} have pointed out that long simulation run times are required to achieve a settled state for the disc from which accurate inferences can be drawn. They show that for their parameters the simulations must be performed for several hundreds of binary orbits for the mass distributions and torques to settle to a quasi-steady configuration. This is consistent with our results for the thin disc case, and we find that the timescale to reach a quasi-steady state is reduced for our thick disc simulations -- as expected as the kinematic viscosity is significantly larger in this case. As discussed in Section~\ref{results} a drawback of the SPH method for simulations of circumbinary discs is that, as the resolution follows the mass, the disc resolution decreases over time as matter accretes on to the binary and the disc expands to larger radii. Thus in the future we will explore simulations in which matter is added to the disc at a suitably large radius, such that the disc resolution can be kept constant over longer timescales. This may provide a decrease in overall computational cost as fewer particles could be utilized to model the initial disc. However, we note that we do not expect the general results (e.g.\@ the binary evolution and inner disc properties) to change from what we have presented here, as we have shown that these properties persist for hundreds of binary orbits and at several different resolutions for the disc. Further, if the direction of the torque depends sensitively on the outer boundary conditions, or running the simulations to a true steady-state\footnote{Note that almost no astrophysical disc is able to achieve a true steady state. The exception is discs formed via mass transfer in binary systems, where the mass supply rate can be very nearly constant for timescales that are much longer than the disc viscous timescale. However, for circumbinary discs it is hard to imagine a mechanism for feeding mass to the circumbinary disc at a constant rate over such long time periods.}, then simulations of isolated discs that are already circular, planar and set up with a power-law surface density profile (i.e. assuming approximate steady-state as an initial condition) are clearly not representative of the problem. Instead detailed knowledge of the real boundary conditions (feeding rate, direction etc) are required for each astrophysical scenario to understand the binary evolution. Fortunately, it appears this is not the case, and the magnitude (and direction) of the torque is determined by the competition between the capture torque (spin up) and resonant torque (spin down), as discussed in Section~\ref{discbinary}, and these are typically determined locally in the disc near the binary.

Finally, a related interesting feature of our simulations concerns the inner disc radius. We begin each simulation with an inner disc edge located at $R_{\rm in} = 3a$ (larger than previous simulations in the literature), as we expected this to be {\it outside} the tidal truncation radius for our disc-binary parameters. In the thick disc simulation, the inner disc edge moves inwards with time. After a time of $\approx 10t_{\rm b}$ the inner edge has moved inwards to approximately $R_{\rm in} \approx 2a$, consistent with expectations, and after $\approx 400t_{\rm b}$ the inner edge has remained in that same location. During this time the surface density of the thick disc is approximately constant from $R=2a$ to $R=5a$, and then falls off steeply for $R<2a$, allowing us to identify $R_{\rm in} \approx 2a$. In contrast the thin disc simulation exhibits a time variable inner disc edge. We caution that there is no obvious precise definition of the disc inner edge in such a dynamic simulation as the azimuthally averaged surface density has no sharp boundary, but we can estimate it as being given approximately by the largest radius at which the surface density strongly and monotonically decays for all smaller radii. Using this loose definition, between 10s of binary orbits and up to $\approx 800t_{\rm b}$ the inner edge oscillates between $\approx 2a$ and $\approx 3a$. It would therefore be interesting to try additional simulations in the future which begin with an inner edge of say $R_{\rm in} = 5a$, and see whether the inner edge moves in and recovers the same behaviour we find here. We conjecture here, that if this is done, or if simulations with a smaller initial inner disc radius are left running for sufficiently long times (and with sufficiently high spatial resolution), that the steady-state solution is one in which the cavity is either full and contains streams of gas flowing on to the binary (thick case) or one in which the cavity is essentially empty and the inner disc regions are essentially circular and relatively featureless at the tidal truncation radius -- i.e.\@ the disc is a decretion disc.

\section{Conclusions}
\label{conclusions}
We have provided physical arguments and supporting numerical simulations that show that the binary orbital evolution resulting from interaction with a circumbinary disc is sensitive to the magnitudes of the resonant torques and viscous torques \citep{Papaloizou:1977aa,Artymowicz:1994aa}.  When the disc is sufficiently thick (high pressure, high viscosity) the capture rate on to the binary becomes sufficient to provide a positive net torque (recall that the angular momentum per unit mass of captured material is larger than the specific angular momentum of the binary as the disc orbits become unstable at a radius greater than the binary orbit, cf. \citealt{Papaloizou:1977aa,Paczynski:1977aa}). However, for the more realistic case where the disc is thinner, with $H/R \lesssim 0.1$, the binary orbital evolution follows the standard paradigm and the binary shrinks with time. This is principally because, in this case, the resonances from the binary are sufficiently strong to impede the accretion flow. And thus while some mass may be transferred from the disc to the binary \citep{Artymowicz:1994aa,Artymowicz:1996aa} the disc more closely resembles a decretion disc than a standard accretion disc. More specifically the disc is subject to a dynamically important central torque, which \cite{Nixon:2020ab} describe through their $f$ parameter which is the ratio of the outward viscous flux of angular momentum from the inner boundary to the inward advected flux of angular momentum there. In the thick disc case we have $f \ll 1$, while in the thin disc case we have $f \gg 1$.

The details of the outcome of disc-binary interactions has important implications for different types of astrophysical systems. Circumbinary discs routinely occur in star-forming regions during chaotic star formation and are thought to occur around supermassive black hole binaries in galactic centres following galaxy mergers. In each of these cases the binary orbital evolution is critical to determining the outcome and for example, whether gas discs can provide a solution to the last parsec problem and efficiently merge SMBH binaries on short timescales. As discussed above it has been concluded in the recent literature, from the results of a small number of numerical simulations that cover only a small range of physical parameters, that circumbinary discs cause the binary to expand. If this were the case, then this would have strong implications for binary orbital evolution and the fate of several distinct astrophysical systems. However, here we have argued (see also \citealt{Tiede:2020aa}) that this conclusion was premature, and that instead, we note here, that the outcome of disc-binary interactions depends in a complex and subtle way on the interplay of several parameters.

\begin{figure}[!htp]
  \includegraphics[width=0.485\textwidth]{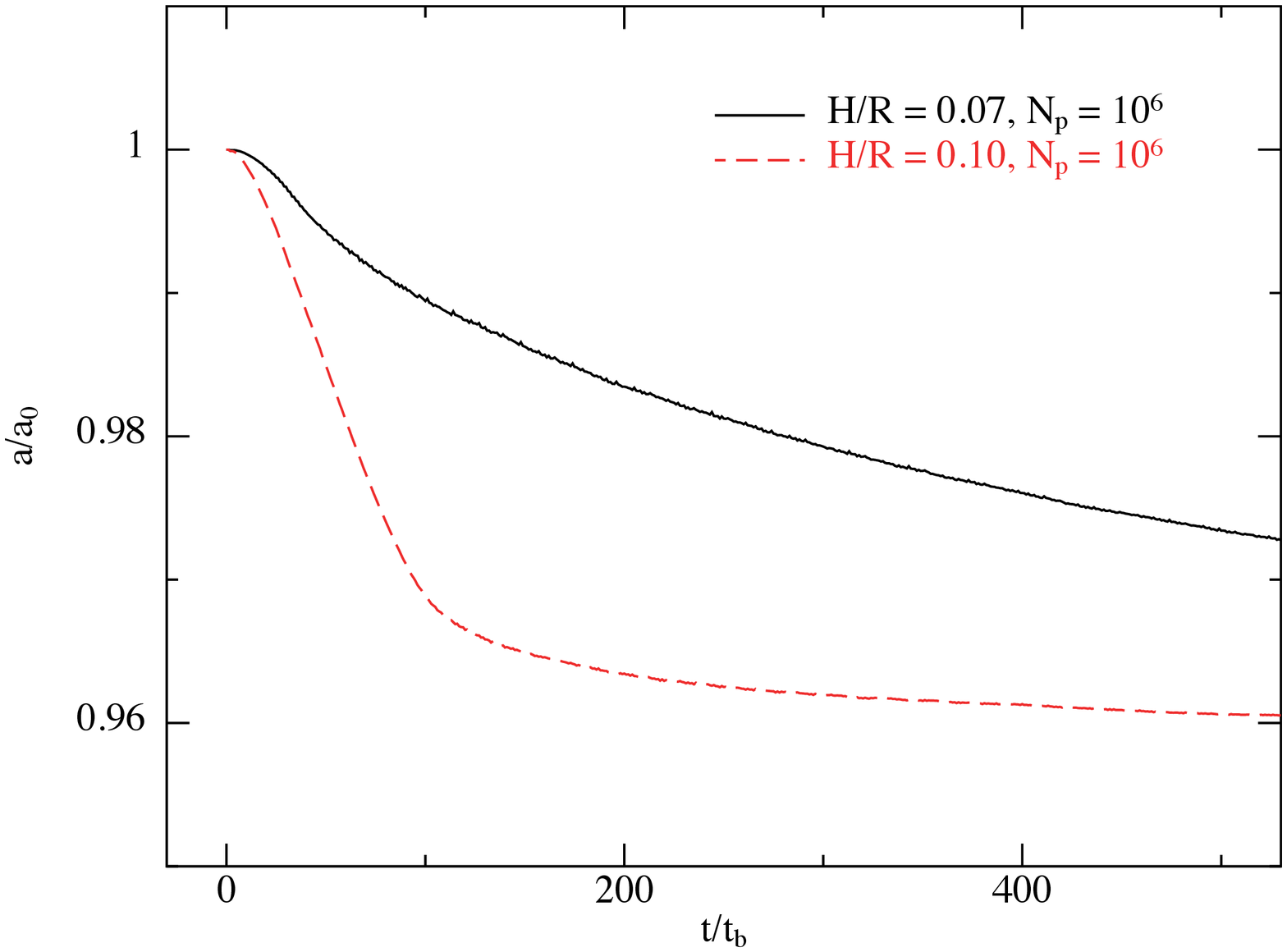}
  \caption{Time evolution of the binary semi-major axis for two additional preliminary simulations with $H/R = 0.07$ (black solid line) and $H/R=0.1$ (red dashed line), both with $\alpha=0.1$ and $N_{\rm p} = 10^6$ (and otherwise the same parameters as the simulations presented in Section~\ref{simulations}). Both of these cases show the binary orbit decaying with time, suggesting that the critical value of $H/R$ that divides binary expansion or contraction is larger than the value of $0.04$ reported by \cite{Tiede:2020aa} when $\alpha \approx 0.1$, and is also parameter dependent (see Section~\ref{discbinary} \& Section~\ref{discussion}). For the simulated parameters, these simulations suggest that the critical $H/R$ is $\approx 0.1-0.2$, as indicated by the small (but still negative) ${\rm d}a/{\rm d}t$ for the $H/R=0.1$ case seen at $t \gtrsim 100t_{\rm b}$.}
  \label{Fig4}
\end{figure}

\cite{Tiede:2020aa} find that the critical $H/R$ value dividing the binary evolution between expansion and contraction lies around $H/R = 0.04$. From the discussion in this paper, it is clear that the dividing line is dependent on other system parameters. We provide in Fig.~\ref{Fig4} some preliminary results of additional simulations we are performing that span a broader range of parameters. In this figure we show the evolution of the semi-major axis of the binary for the same parameters described in Section~\ref{simulations}, but with $\alpha = 0.1$ and two intermediate values of the disc thickness, $H/R = 0.07$ and $H/R = 0.1$. These preliminary results are for discs with $N_{\rm p} = 10^6$, and therefore we can be reasonably confident that the evolution is correct (cf. Fig.~\ref{Fig2}), but cannot provide strong conclusions until higher resolution simulations have also completed, which we leave to a subsequent paper. However, Fig~\ref{Fig4} shows that in both cases with intermediate $H/R$ the binary shrinks with time. This suggests that the critical value of $H/R$ (for near-circular, equal-mass binaries, with $\alpha \approx 0.1$) is higher than the value reported by \cite{Tiede:2020aa} and closer to $\approx 0.1-0.2$. Therefore, from the arguments we have provided, and the numerical simulations we have presented, we speculate that physical expansion of the binary orbit in this case is limited to a small region of parameter space in which the disc is sufficiently thick to weaken the torque applied to the disc and sufficiently viscous to enforce strong mass flow rates on to the binary. We therefore expect that many astrophysical binary systems, which do not possess such extreme parameters, shrink with time upon interacting with a circumbinary disc. For example, if the structure of circumbinary discs around SMBH binaries resembles the structure of accretion discs in AGN \citep[e.g.][]{Collin-Souffrin:1990aa}, then it is expected that most SMBH binaries shrink with time while interacting with circumbinary discs as typically $H/R \approx 2\times10^{-3}$ in this case.

There may be some instances where accretion discs may be thick enough to cause a binary to expand, and these typically occur where the temperature is set by a parent object and the gravitational potential set by a daughter object, such as a circumplanetary disc that is heated by the parent star \citep[cf.][]{Martin:2011ab} or for stellar mass black hole binaries orbiting inside AGN discs \citep[cf.][and references therein]{McKernan:2018aa}. In this latter case, assuming that the AGN disc has $H/R \approx 2\times 10^{-3}$ across a broad range of radii \citep[see][]{Collin-Souffrin:1990aa}, and that the circumbinary disc formed around an embedded stellar-mass black hole binary (of mass $M_{\rm b}$ and orbital distance $R_{\rm b}$ from the central SMBH of mass $M_{\rm h}$) has radial size of order $R_{\rm d} \sim 0.3R_{\rm H}$, where $R_{\rm H} = R_{\rm b}(M_{\rm b}/M_{\rm h})^{1/3}$ is the stellar-mass binary's Hills sphere, then, if the disc thickness is comparable between the AGN disc and the circum-stellar-mass-black-hole-binary-disc, we have $H/R_{\rm d} \sim 0.2\mu^{-1/3}$ where $\mu = 5\times 10^{-5} M_{\rm h}/M_{\rm b}$. Thus the orbital evolution of embedded stellar binary systems in AGN discs is at best unclear.

An additional scenario in which a binary might expand due to interaction with a circumbinary disc is a wide-separation (greater than 10s of au) protostellar binary, where the circumbinary disc is passive and heated by the protostellar light. In this case, the standard model for a circumstellar disc presented by \cite{Chiang:1997aa} has large disc thicknesses that approach unity at radii $\approx 200$\,au (see their equation 14). If a protostellar binary was surrounded by a circumbinary disc with these properties and on these scales, then expansion of the binary orbit may be possible. However, the disc viscosity in these discs and on these scales is a matter of debate, and thus the outcome remains unclear.

\begin{acknowledgements}
We thank Jim Pringle for useful discussions, and the referee for a useful report. We thank Christopher Tiede for sending a copy of \cite{Tiede:2020aa} prior to its publication. CJN is supported by the Science and Technology Facilities Council (grant number ST/M005917/1). This project has received funding from the European Union’s Horizon 2020 research and innovation program under the Marie Sk\l{}odowska-Curie grant agreement No 823823 (Dustbusters RISE project). This research used the ALICE High Performance Computing Facility at the University of Leicester. This work was performed using the DiRAC Data Intensive service at Leicester, operated by the University of Leicester IT Services, which forms part of the STFC DiRAC HPC Facility (\url{www.dirac.ac.uk}). The equipment was funded by BEIS capital funding via STFC capital grants ST/K000373/1 and ST/R002363/1 and STFC DiRAC Operations grant ST/R001014/1. DiRAC is part of the National e-Infrastructure. We used {\sc splash} \citep{Price:2007aa} for the figures.
\end{acknowledgements}

\bibliographystyle{aasjournal}
\bibliography{nixon}
\end{document}